\documentclass[11pt,superscriptaddress,nofootinbib]{revtex4}
\usepackage{graphicx}
\usepackage{amssymb}
\usepackage{amsmath}

\newcommand{\be}{\begin{eqnarray}}
\newcommand{\ee}{\end{eqnarray}}
\newcommand{\ra}{\rightarrow}
\newcommand{\D}{\mathrm{d}}

\newcommand{\E}{\mathrm{e}}

\newcommand{\mL}{\mathcal{L}}
\newcommand{\mK}{\mathcal{K}}

\begin{document}

\title{Stick-slip motion of solids with dry friction subject to random vibrations and an external field}

\author{A. Baule}
\affiliation{The Rockefeller University, 1230 York Avenue, New York, NY 10065, USA}

\author{H. Touchette}
\affiliation{School of Mathematical Sciences, Queen Mary University of London, London E1 4NS, UK}

\author{E. G. D. Cohen}
\affiliation{The Rockefeller University, 1230 York Avenue, New York, NY 10065, USA}

\begin{abstract}
We investigate a model for the dynamics of a solid object, which moves over a randomly vibrating solid surface and is subject to a constant external force. The dry friction between the two solids is modeled phenomenologically as being proportional to the sign of the object's velocity relative to the surface, and therefore shows a discontinuity at zero velocity. Using a path integral approach, we derive analytical expressions for the transition probability of the object's velocity and the stationary distribution of the work done on the object due to the external force. From the latter distribution, we also derive a fluctuation relation for the mechanical work fluctuations, which incorporates the effect of the dry friction.
\end{abstract}


\date{\today}

\maketitle

\section{Introduction}

This paper is a sequel to a previous paper \cite{Baule10}, which studied the motion of a solid object (e.g., a coin) moving with friction over a horizontal, uniformly rough plate, which is vibrated laterally by external Gaussian white noise. Here, we extend this study by adding a constant external field, so that we effectively consider the object's motion on an inclined plate with the added influence of gravity. The evolution of the velocity $v(t)$ of the object relative to the plate is taken to be given by the Langevin equation
\begin{eqnarray}
\label{model}
m\dot{v}(t)+\alpha v(t)+\sigma(v(t))\Delta_F=F+\xi(t),
\end{eqnarray}
where $F$ is the external force associated with gravity and $\xi(t)$ is an external noise term associated with the random acceleration of the plate. As in \cite{Baule10}, we assume that this noise is a Gaussian white noise characterized by 
\be
\left<\xi(t)\right>=0,\qquad
\left<\xi(t)\xi(t')\right>=2D\delta(t-t'),
\label{noise_corr}
\ee
where the brackets indicate an average over the noise, and $D$ is the noise strength.

The friction between the moving object and the plate is phenomenologically described in Eq.~(\ref{model}) by two friction coefficients: a (fluid-like) dynamical friction coefficient $-\alpha v$, with strength $\alpha$, which is the solid analogue of the viscous (Stokes) friction in fluids, and a static (dry) solid-solid friction $-\sigma(v)\Delta_F$, where $\Delta_F$ is the strength of the dry friction and the sign function $\sigma(v)$ is defined as equal to $+1$, $0$, $-1$ for $v>0$, $v=0$, $v<0$, respectively. The latter force is often referred to as Coulomb friction,\footnote{The history of solid-solid or dry friction goes back a long time; see, e.g., \cite{Persson} and references therein.} 
and was studied in the context of Brownian motion by de Gennes \cite{deGennes05}. The main feature of this force, which is singular at $v=0$, is that it can lead to the sticking of the object on the plate when $v=0$ and the total external force $F+\xi$ in Eq.~(\ref{model}) is smaller in magnitude than the ``contact'' force $\Delta_F$. When $v\neq 0$ or $|F+\xi|>\Delta_F$, then the dry friction simply adds to the viscous friction force, and so contributes to the sliding motion of the object (slip motion).

Since the object is under the influence of two types of external forces, namely the random vibrations of the underlying plate, $\xi(t)$, and the external force, $F$, it can be considered to be in a nonequilibrium state. In the following, we will investigate the statistical properties of this state by obtaining analytic expressions for the transition probability or propagator $f(v,t|v_0,t_0)$, which gives the probability that the object has a velocity $v$ at time $t$ starting from an initial velocity $v_0$ at time $t_0<t$. The propagator is obtained using path integral techniques in the low-noise limit $D\rightarrow 0$, and yields, in the limit $t-t_0\rightarrow\infty$, a stationary velocity distribution $p_s(v)$ \cite{Hayakawa05}, which characterizes the nonequilibrium steady state (NESS) of the object, reached when it moves for a long time compared with its intrinsic relaxation time. In this limit, we also obtain the large deviation approximation of the probability distribution $P(W_\tau)$ of the mechanical work $W_\tau$ performed by the external force $F$ on the object during the time interval $[t,t+\tau]$. This work $W_\tau$ is defined as
\be
\label{work}
W_\tau[v(s)]=F\int_{t}^{t+\tau} v(s)\D s,
\ee
and is obviously a random variable, since it is a functional of the random velocity $v(s)$. From the large deviation approximation of $P(W_\tau)$, we then obtain a so-called fluctuation relation for the work fluctuations, which characterizes the combined effect of the dry friction and the fluid-like friction in the NESS. 

The motion of two solids over each other with friction is a ubiquitous problem in nature, of great practical and theoretical importance, which occurs, e.g., in geology, physics, engineering and biology. Compared with viscous or fluid friction, which is relatively well understood from the microscopic point of view, dry friction is quite a complicated process, which may be modeled at different scales using different physical scenarios (e.g., plastic deformations, breaking of microscopic bonds, etc.~\cite{Persson}). The term $-\Delta_F\sigma(v)$ should obviously be considered only as a phenomenological or effective model of dry friction. It is a very simplified model when compared to naturally- or technologically-occurring situations, but one which incorporates some of the basic features of dry friction, in particular, the possibility of alternating stick-slip motion. From the point of view of nonequilibrium systems, the model of Eq.~(\ref{model}) can also be thought of as a simple phenomenological model with nonlinear friction, for which the properties of the work fluctuations in a NESS can be investigated theoretically and experimentally. We comment on possible experimental realizations of this model in the concluding section in the paper.

\section{Preliminaries}

The mathematical technique that we will use to obtain the transition probability $f(v,t|v_0,t_0)$ of Eq.~(\ref{model}) was already described in \cite{Baule10}. Here we briefly repeat the basis of this technique, and recall a few properties of the deterministic equation describing the motion of the object in the absence of noise, i.e., $\xi(t)=0$. These properties will be useful for understanding all the steps involved in the derivation of analytical expressions for $f(v,t|v_0,t_0)$. 

\subsection{Path integral formalism}

In order to obtain the transition probability $f(v,t|v_0,t_0)$, we use a path integral approach in which $f(v,t|v_0,t_0)$ is formally expressed as an integral over all paths $v(s)$ leading from an initial state $(v_0,t_0)$ to a final state $(v,t)$. To write this path integral, it is convenient to divide Eq.~(\ref{model}) by the mass $m$ and put the resulting equation in the form
\be
\label{bm}
\dot{v}(t)=-U'(v(t))+\xi(t)/m,
\ee
where
\be
\label{U_pot}
U(v)=\frac{v^2}{2\tau_m}+|v|\Delta-a\,v,
\ee
is the effective potential asssociated with the friction forces and the constant external force present in the Langevin equation, $\tau_m= m/\alpha$ is the characteristic inertial time scale, $\Delta=\Delta_F/m$ is the threshold acceleration associated with the dry friction force, and $a= F/m$ is the acceleration associated with the external force. In terms of $U(v)$, the stationary distribution of Eq.~(\ref{bm}), obtained in the asymptotic limit $t\rightarrow\infty$, takes a Boltzmann-like form
\be
\label{p_stat}
p_s(v)= Ne^{-\gamma U(v)},
\ee
where $\gamma= m^2/D$ and $N$ is a normalization constant \cite{Risken}. As for $f(v,t|v_0,t_0)$, it is expressed as \cite{Feynman}
\begin{eqnarray}
\label{path_int}
f(v,t|v_0,t_0)=\int_{(v_0,t_0)}^{(v,t)} J[v(s)]e^{-\gamma A[\dot{v}(s),v(s)]}\mathcal{D}v(s),
\end{eqnarray}
where $A$ is a functional of $v(s)$,
\begin{eqnarray}
\label{action}
A[\dot{v}(s),v(s)]=\int_{t_0}^t\mathcal{L}(\dot{v}(s),v(s))\D s,
\end{eqnarray}
which will be referred to as the \textit{action} associated with the path $v(s)$, and $\mathcal{L}$ is an Onsager-Machlup Lagrangian \cite{Onsager53,Machlup53} given by 
\begin{eqnarray}
\label{lagrangian}
\mathcal{L}(\dot{v}(s),v(s))=\frac{1}{4}\left[\dot{v}(s)+U'(v(s))\right]^2.
\end{eqnarray}
In Eq.~(\ref{path_int}), the integral $\int \mathcal{D}v(s)$ denotes an integral over all paths $v(s)$ from $(v_0,t_0)$ to $(v,t)$. The Jacobian $J[v(s)]$ originates from the transformation $\xi(t)\rightarrow v(t)$ in Eq.~(\ref{bm}) and is a functional of $v(s)$ \cite{Graham73,Hunt81}:
\be
\label{jacobian}
J[v(s)]=e^{\frac{1}{2}\int_{t_0}^t U''(v(s))\D s}.
\ee

Following our previous work \cite{Baule10}, we will evaluate the path integral (\ref{path_int}) in the limit $\gamma\ra \infty$, using the saddlepoint approximation. In this limit, the dominant contribution to the path integral is due to a particular path $v^*(s)$ which maximizes the exponent in Eq.~(\ref{path_int}), or equivalently, which minimizes the action $A$. As a result, $v^*(s)$ must satisfy the Euler-Lagrange (E-L) equation
\be
\label{ELform}
\frac{\D}{\D s}\frac{\partial \mathcal{L}}{\partial \dot{v}^*}-\frac{\partial \mathcal{L}}{\partial v^*}=0.
\ee
The path $v^*(s)$ corresponds to the path with the highest probability among all paths connecting $(v_0,t_0)$ and $(v,t)$, and is called, for this reason, the \textit{optimal} path or simply the most probable path connecting $(v_0,t_0)$ and $(v,t)$.\footnote{In the following, we indicate optimal paths $v^*(s)$ with the superscript $*$ and drop the argument $s$ whenever possible without confusion, so that then $v^*(s)$ is just abbreviated to $v^*$.} Once this path is found, the transition probability is approximated as
\be
\label{sp_approx}
f(v,t|v_0,t_0)\approx e^{-\gamma A[\dot{v}^*,v^*]},
\ee
where we neglect the contribution of the Jacobian as $\gamma\ra \infty$.\footnote{The Jacobian may be significant for finite $\gamma$ and has then to be taken into account, e.g., when comparing the saddlepoint approximation Eq.~(\ref{sp_approx}) with numerical results.}

The next sections present the solution of the E-L equation in full detail. When choosing initial conditions for this equation, it is important to note that, without $a$, the dynamics described by Eq.~(\ref{bm}) exhibits a statistical forward/backward symmetry, in the sense that the probability of observing the object's velocity $v$ at time $t$ for a given initial velocity $v_0$ at $t_0$ is identical to that of observing $-v$ at $t$ for the initial velocity $-v_0$ at $t_0$ \cite{Baule10}. The presence of the constant external acceleration $a$ breaks this symmetry, and so one has to distinguish between positive and negative $v_0$ for a given $a$. However, the symmetry is recovered upon changing both $v_0\ra -v_0$ and $a\ra -a$, so that in the following, we will only consider $a>0$ without loss of generality, and will distinguish between the two cases $v_0>0$ and $v_0<0$.

\subsection{Deterministic motion}

\begin{figure}
\begin{center}
\begin{tabular}{ll}
\includegraphics[width=7.5cm]{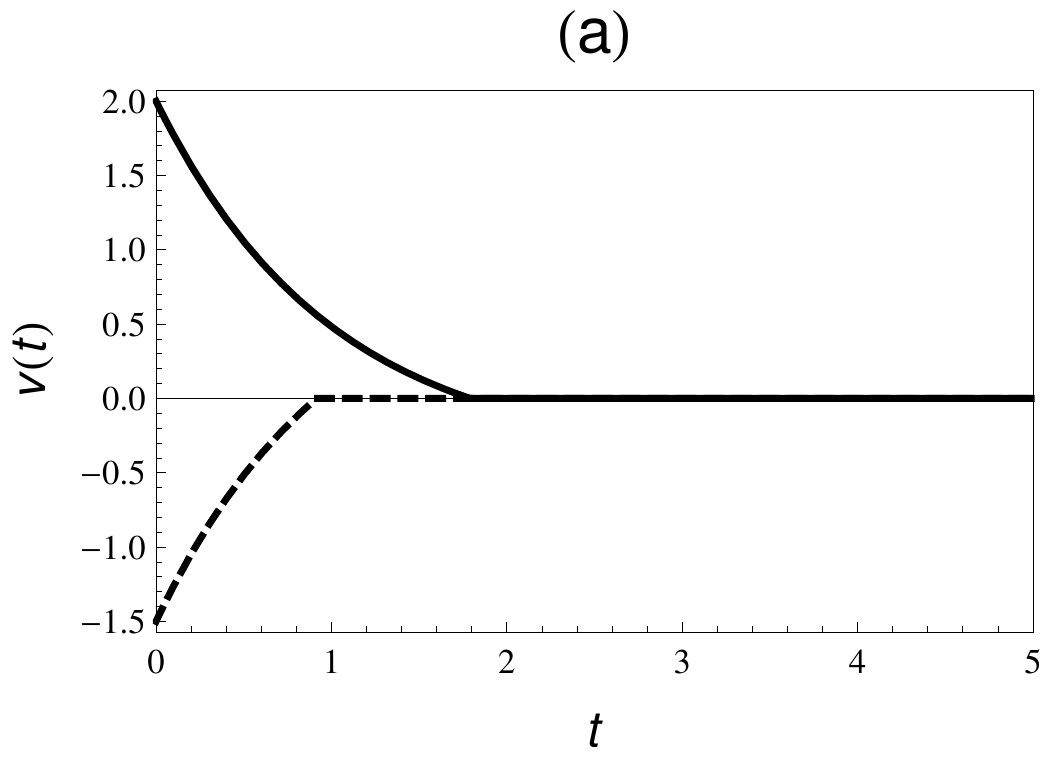} &\hspace{0.5cm}\includegraphics[width=7.5cm]{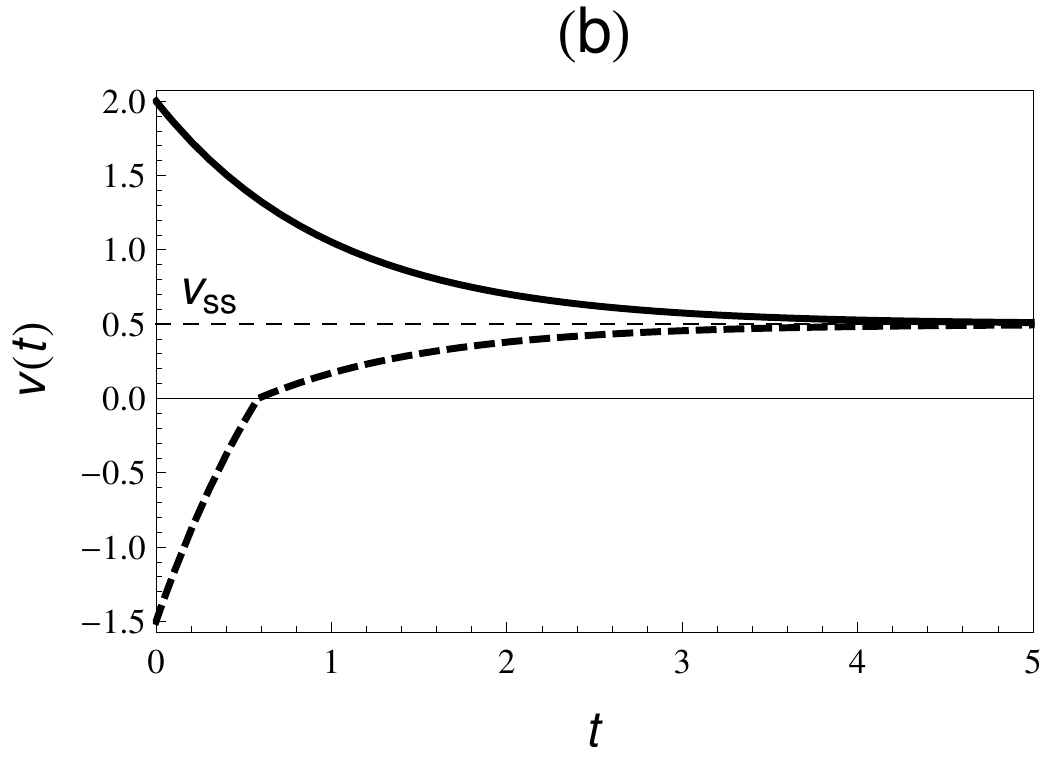}
\end{tabular}
\caption{\label{Fig_traj_F}Two examples of the object's velocity $v(t)$ described by Eq.~(\ref{det_eq}). (a) The case $a<\Delta$. For both positive (solid curve) and negative (solid dashed curve) initial velocities, the object gets stuck after a finite time $t_{st}$, i.e., $v(t)=0$ for all $t\geq t_{st}$. At $t_{st}$, $\dot{v}$ is discontinuous due to the $\sigma(v)$ singularity in Eq.~(\ref{det_eq}). (b) The case $a>\Delta$. For both positive (solid curve) and negative (solid dashed curve) initial velocities, the object approaches the constant velocity $v_{ss}$, Eq.~(\ref{v_st}), (thin dashed line) for times larger than the relaxation time $\tau_m$. When crossing the $v=0$ axis, the slope $\dot{v}(t)$ is discontinuous. Parameter values: $\tau_m=1.0$, $\Delta=0.7$. (a) $a=0.3$. (b) $a=1.2$.}
\end{center}
\end{figure}

Before we treat the random motion given by the Langevin equation (\ref{bm}), it is instructive to consider the deterministic version of that equation obtained by setting $\xi(t)=0$:
\be
\label{det_eq}
\dot{v}(t)&=&-\frac{1}{\tau_m}v(t)-\sigma(v(t))\Delta+a.
\ee
Two types of motion arise when solving this equation:

(a) Stick: When $a<\Delta$, for both $v(t)>0$ and $v(t)<0$, the combined acceleration on the right hand side of Eq.~(\ref{det_eq}) always acts opposite to the motion, so that the $v(t)=0$ axis is an attractor of the dynamics (cf.~Fig.~\ref{Fig_traj_F}(a)). This implies that the deterministic path effectively follows the $v=0$ axis, even though $v(t)=0$ is formally \textit{not} a solution of Eq.~(\ref{det_eq}). Physically, this behavior is expected: if the vibrating plate is sufficiently rough and the external force sufficiently weak, the object should get stuck, i.e., reach a zero velocity for any initial velocity $v_0$. 

(b) Slip: When $a>\Delta$ on the other hand, the combined acceleration due to the dry friction and the external force, $-\sigma(v)\Delta+a$ in Eq.~(\ref{det_eq}), is always positive. This means that in this case the object eventually moves away from the $v=0$ axis and approaches then a constant non-zero velocity attractor $v_{ss}$ (cf.~Fig.~\ref{Fig_traj_F}(b)), which is given by
\be
\label{v_st}
v_{ss}=(a-\Delta)\tau_m.
\ee
Importantly, the approach to the zero velocity attractor of the motion is in a finite time, as in the force free $a=0$ case (cf.~Fig.~\ref{Fig_traj_F}(a) and \cite{Baule10}), while the non-zero attractor is approached with an exponential decay over an infinite time.

In the following, we will encounter many situations similar to case (a) above where the motion of the object is well defined for $v>0$ or $v<0$, but not at $v=0$ because of the singularity of $\sigma(v)$. The procedure that we shall adopt to treat such situations is to invoke continuity: physically, the velocity of the object cannot jump in time, so that $v(s)$ must be a continuous function of $s$. In practice, this means that the motion of the object should be determined by considering its equation of motion in the regions $v>0$ and $v<0$ separately, and by matching at the origin the solutions obtained, if needed. Figure~\ref{Fig_traj_F}(b) shows an example of a solution of Eq.~(\ref{det_eq}) obtained from this procedure when $v_0<0$ and $a>\Delta$. In this case, the solution must physically cross the $v=0$ axis in a continuous way to reach the positive steady-state velocity $v_{ss}$, given by Eq.~(\ref{v_st}). As the solution goes from the lower to the upper half of the $(v,s)$-plane, it changes slope discontinuously because of the sign change of the dry friction term. Such a discontinuity in the slope of $v(s)$ will often be encountered in what follows.

\section{Optimal paths in the velocity-time plane}

\begin{figure}
\begin{center}
\begin{tabular}{ll}
\includegraphics[width=7.5cm]{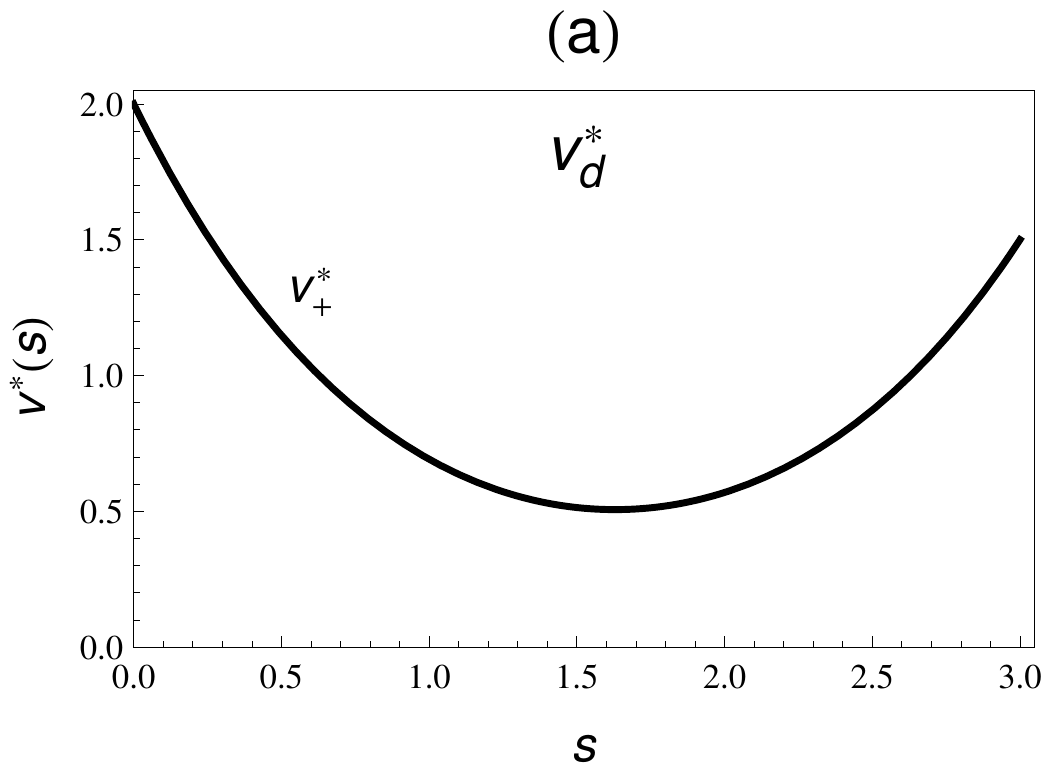} &\hspace{0.5cm}\includegraphics[width=7.5cm]{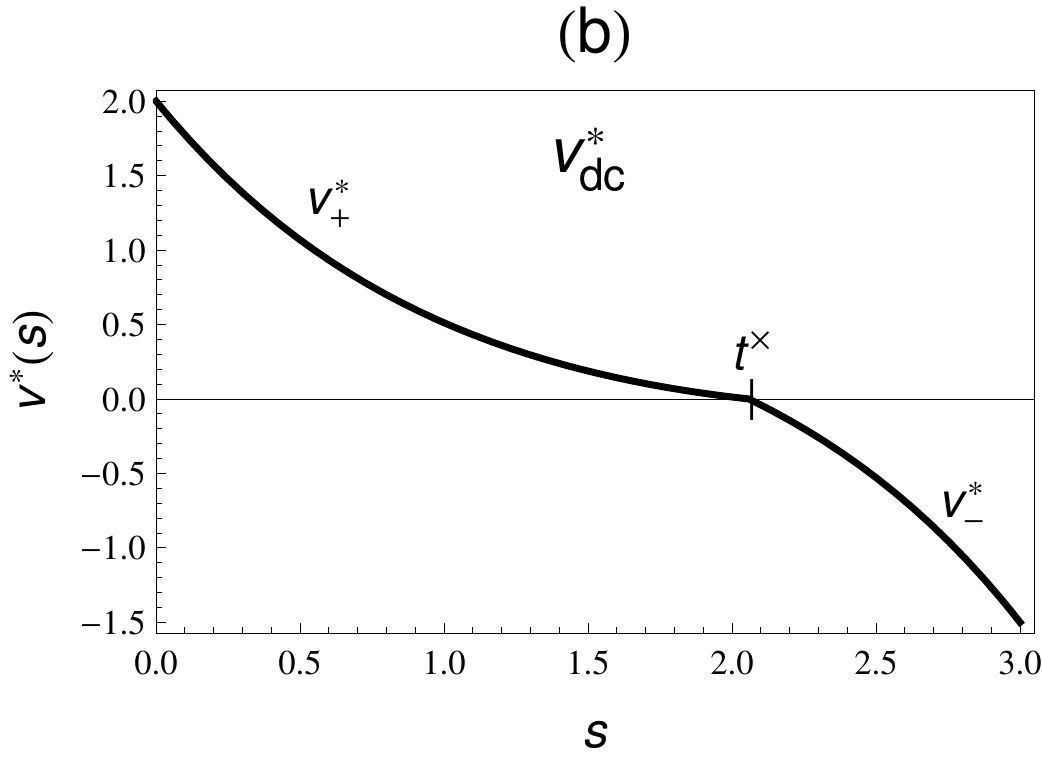} \vspace{0.5cm}
\\
\includegraphics[width=7.5cm]{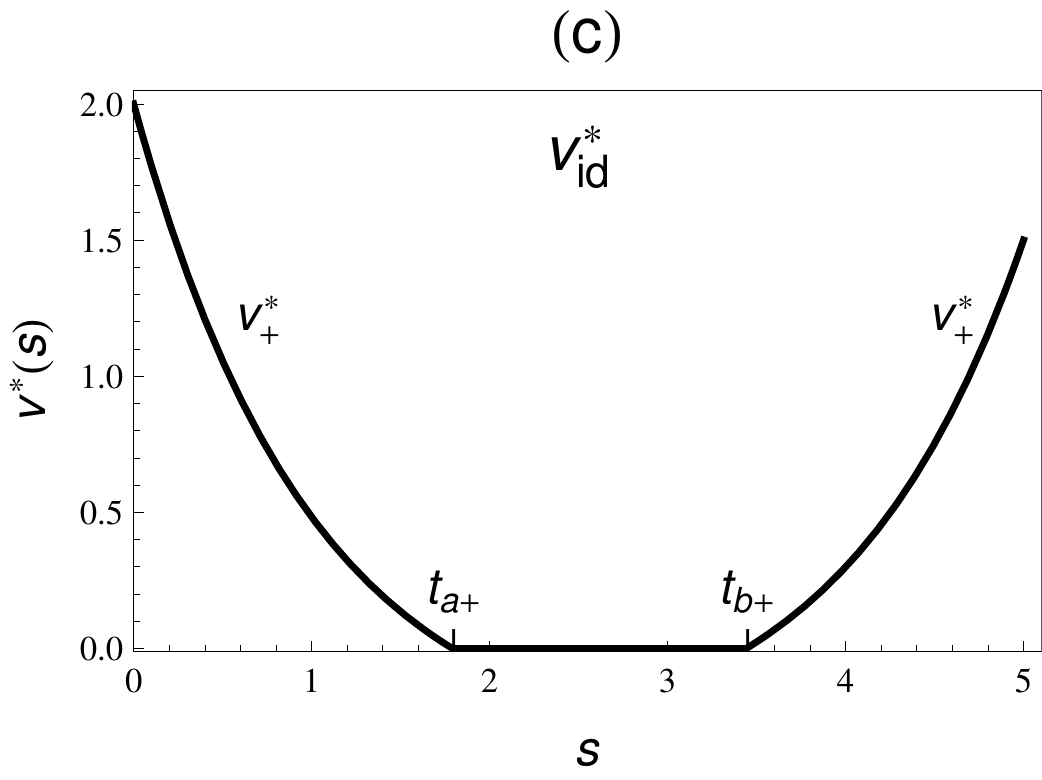} &\hspace{0.5cm}  \includegraphics[width=7.5cm]{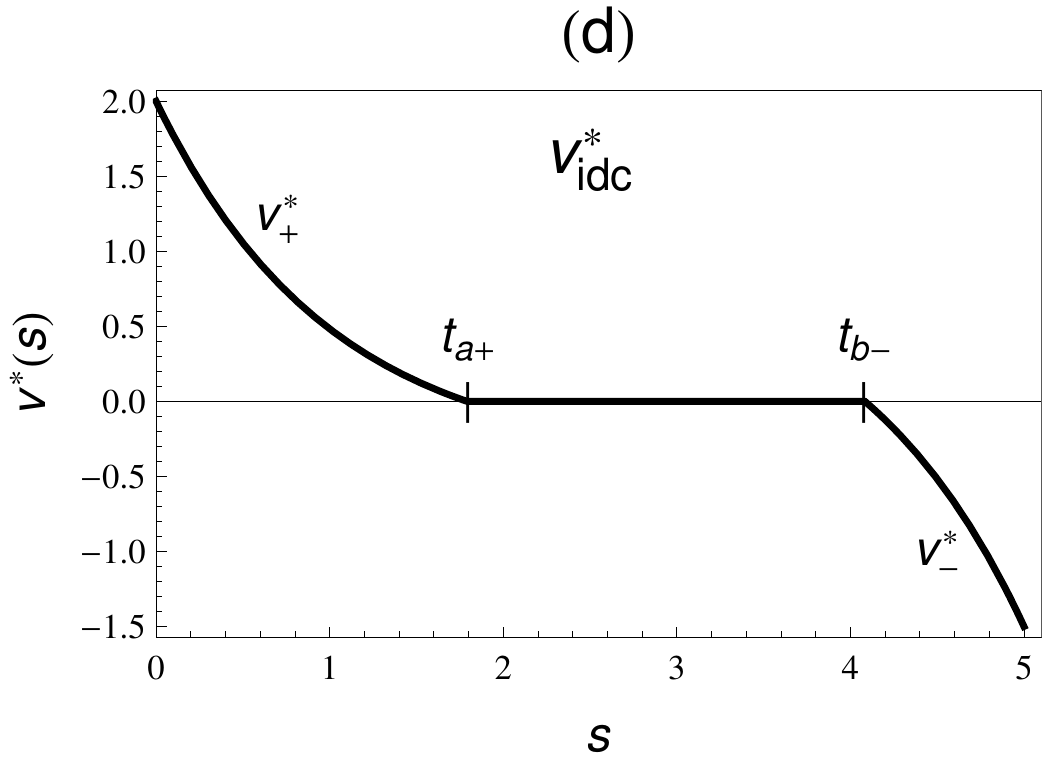}
\end{tabular}
\caption{\label{Fig_smallF}Examples of optimal paths $v^*(s)$ in the case $a<\Delta$ for $v_0>0$. The behavior for $v_0<0$ follows by reflection of the paths at the $v^*=0$ axis. (a) Direct path on the upper $(+)$ half plane. (b) Direct crossing path on two half planes, where the crossing time is denoted by $t^\times$. At $v^*=0$ the curvature changes from a positive to a negative value and the derivative $\dot{v}^*$ is discontinuous. (c), (d) Indirect paths: for sufficiently large times the optimal path follows the attractor $v^*=0$ for a finite time. At the crossovers to/from $v^*=0$ at the times $t_{a+}$ and $t_{b\pm}$, respectively, the slope $\dot{v}^*$ is discontinuous. The $\pm$ on the times $t_a$ and $t_b$ refer to the upper and lower half planes, on which the paths are, respectively. Parameter values: $\tau_m=1.0$, $\Delta=0.7$, $a=0.3$.}
\end{center}
\end{figure}

We present in this section the solution of the E-L equation (\ref{ELform}) associated with the Lagrangian (\ref{lagrangian}). Following the previous section, we write the E-L equation for the optimal path $v^*(s)$ as two equations:
\begin{subequations}
\label{EL}
\begin{align}
\ddot{v}^*-\frac{1}{\tau_m^2}v^*-\frac{1}{\tau_m}(\Delta-a)&=0\qquad v^*>0
\label{EL1}\\
\ddot{v}^*-\frac{1}{\tau_m^2}v^*+\frac{1}{\tau_m}(\Delta+a)&=0\qquad v^*<0,\label{EL2}
\end{align}
\end{subequations}
which are supplemented by the requirement that $v(t)$ be continuous across the boundary $v=0$. To treat this condition explicitly, we distinguish next between optimal paths which always remain on one one half plane and optimal paths which cross the singular $v^*=0$ axis to span the two half planes:

(a) If the boundary conditions of the E-L equation are such that the optimal path remains entirely either on the upper ($v^*>0$) or on the lower half ($v^*<0$) of the velocity-time plane, respectively, one finds the optimal paths simply by solving either Eq.~(\ref{EL1}) or Eq.~(\ref{EL2}). The two solutions are
\be
\label{sol_basic}
v^*_\pm(s)=B_\pm e^{s/\tau_m}+C_\pm e^{-s/\tau_m}+(a\mp\Delta)\tau_m,
\ee
where the subscripts $+$ and $-$ refer to the upper and lower half plane, respectively. The prefactors $B_\pm$ and $C_\pm$ are determined from the boundary conditions
\be
v^*_\pm(t_0)=v_0, \qquad v^*_\pm(t)=v
\ee
and are straightforwardly given by
\be
\label{B+}
B_\pm&=&\frac{e^{t/\tau_m}(v+(\Delta\mp a) \tau_m)-e^{t_0/\tau_m}(v_0+(\Delta\mp a) \tau_m)}{e^{2t/\tau_m}-e^{2t_0/\tau_m}},\\
\label{C+}
C_\pm&=&\frac{e^{t/\tau_m}(v_0+(\Delta\mp a) \tau_m)-e^{t_0/\tau_m}(v+(\Delta\mp a) \tau_m)}{e^{(t-t_0)/\tau_m}-e^{-(t-t_0)/\tau_m}}.
\ee
Since the basic solutions $v^*_\pm$ are always away from the $v^*=0$ axis, i.e., never have a zero velocity, they represent physically a pure slip motion of the object. We denote such optimal paths as \textit{direct paths on one half plane} (cf.~Fig.~\ref{Fig_smallF}a).

(b) If the boundary conditions are such that the optimal path has to traverse the singularity at the $v^*=0$ axis (e.g., if the initial velocity is positive and the final velocity negative), $v^*$ will consist of the basic solutions $v^*_+$ and $v^*_-$, Eq.~(\ref{sol_basic}), on the upper and lower half of the $(v^*,s)$-planes, respectively. In addition, at the singular $v^*=0$ axis, where the optimal path crosses from the upper to the lower half plane or vice versa, we impose the condition that the optimal path is continuous, since on physical grounds the object's velocity should be continuous.

Two types of behavior at the singularity $v=0$ are then possible:
\begin{enumerate}
\item The optimal path crosses the $v=0$ axis only at a \textit{single point} in time (cf.~Fig.~\ref{Fig_smallF}b). Optimal paths with a single crossing point are denoted as \textit{direct crossing paths} and correspond physically to a pure slip motion of the object just as direct paths.
\item The optimal path spends a \textit{finite time} at $v^*=0$, i.e., it reaches the $v^*=0$ axis from the initial point $(v_0,t_0)$, and stays on this axis for a finite time, before reaching the final point $(v,t)$ (cf.~Figs.~\ref{Fig_smallF}c and \ref{Fig_smallF}d). This behavior is only possible for $a<\Delta$, because in that case the $v^*=0$ axis is an attractor of the deterministic motion (see the discussion of Eq.~(\ref{det_eq}) above). Since the deterministic dynamics always has zero action, a solution of the deterministic equation also has to represent an optimal path, because the optimal paths are those paths with a minimal action. Therefore, we have to admit optimal paths that follow the $v=0$ axis for a finite time. These optimal paths are denoted as \textit{indirect paths} and correspond physically to a stick-slip motion.
\end{enumerate}

The behavior of these different types of optimal paths is described quantitatively in the next two subsections for the cases $a<\Delta$ and $a>\Delta$, respectively. For this discussion, we first require the actions associated with the basic solutions $v^*_\pm(s)$. Substituting Eq.~(\ref{sol_basic}) into Eqs.~(\ref{lagrangian}) and~(\ref{action}) yields for the action for paths in the upper half plane and lower half, respectively:
\be
\label{lambda}
A[\dot{v}^*_\pm,v^*_\pm]&=& \frac{\left[e^{t/\tau_m}((\Delta\mp a)\tau_m+|v|)-e^{t_0/\tau_m}((\Delta\mp a)\tau_m+|v_0|)\right]^2}{2\tau_m\left(e^{2t/\tau_m}-e^{2t_0/\tau_m}\right)}\nonumber\\
&=&\Lambda_\pm(v,t;v_0,t_0).
\ee
Due to the presence of the constant external acceleration $a$, the action of the optimal paths $v^*_\pm$ are different on the upper and lower half planes.

\subsection{The case $a<\Delta$}
\label{Sec_asmallerD}

In this case, direct paths as well as indirect paths appear for $v_0>0$ and $v_0<0$ (cf.~Fig.~\ref{Fig_smallF}). The \textit{direct} paths on one half plane will be labeled by a subscript $d$ and can be represented by (cf.~Fig.~\ref{Fig_smallF}a) 
\be
\label{dir_par}
\left.v^*_d(s)\right|_{v_0,t_0}^{v,t}=\left.v^*_\pm(s)\right|_{v_0,t_0}^{v,t},
\ee
where the $v^*_\pm$ are given by Eq.~(\ref{sol_basic}) and the boundary conditions $v_0, t_0$ and $v,t$ are explicitly indicated. The subscripts $+$ and $-$ refer here to direct paths ($d$) on the upper and lower half plane, respectively. These direct paths have the associated action
\be
\label{Adir}
A[\dot{v}^*_d,v^*_d]=\Lambda_\pm(v,t;v_0,t_0),
\ee
which follows immediately from Eqs.~(\ref{dir_par}) and (\ref{lambda}).

The \textit{direct crossing} paths will be labeled by a subscript $dc$ and can be represented by (cf.~Fig.~\ref{Fig_smallF}b) 
\be
\label{dirx_par}
v^*_{dc}(s)=
\left\{
\begin{array}{lll} 
\left.v^*_\pm(s)\right|_{v_0,t_0}^{0,t^\times} & &  t_0\le s\le t^\times\\ 
\left.v^*_\mp(s)\right|_{0,t^\times}^{v,t} & & t^\times< s \le t.
\end{array}\right.
\ee
In Eq.~(\ref{dirx_par}), the time when the direct crossing path crosses the $v^*=0$ axis is indicated by $t^\times$, which must satisfy $t_0<t^\times<t$ and can be found from the principle of minimal action. The action of the direct crossing paths is
\be
\label{Adirx}
A[\dot{v}^*_{dc},v^*_{dc}]&=&\Lambda_\pm(0,t^\times;v_0,t_0)+\Lambda_\mp(v,t;0,t^\times),
\ee
which follows from Eq.~(\ref{action}) with Eqs.~(\ref{dirx_par}) and (\ref{lambda}). Thus, the crossing time $t^\times$ is given by the two equations
\be
\label{tbar_eq}
\frac{\partial}{\partial t^\times}[\Lambda_\pm(0,t^\times;v_0,t_0)+\Lambda_\mp(v,t;0,t^\times)]=0,
\ee
which each have a unique real root $t_0<t^\times<t$. Note that throughout this paper the functions $\Lambda_\pm(v,t;v_0,t_0)$ are adapted to the case at hand by replacing $v$, $t$, $v_0$, $t_0$ by the appropriate velocities and times. At the crossover time $t^\times$ the derivative of the optimal path is discontinuous (cf.~Figs.~\ref{Fig_smallF}a and \ref{Fig_smallF}b).

The E-L equations~(\ref{EL}) imply that for $a<\Delta$ the curvature of the optimal path changes from a positive to a negative value upon crossing the $v^*=0$ axis. This implies that the optimal path is then always convex or concave on the upper or lower half plane, respectively. We will see later that this change in curvature leads to \textit{forbidden regions} as in the force free case \cite{Baule10}, in the sense that there exist regions in the $(v^*,s)$-plane that cannot be reached by any direct path from a given initial velocity. Final velocities inside the forbidden region can only be reached by indirect paths. This is discussed in detail in Sec.~\ref{Sec_structure}.

\textit{Indirect} paths consist of three parts: a first part from the initial point $(v_0,t_0)$ to the $v^*=0$ axis at $(0,t_{a\pm})$, a second part along the zero axis from $(0,t_{a\pm})$ to $(0,t_{b\pm})$, and a third part from $(0,t_{b\pm})$ to the final point $(v,t)$ (cf.~Figs.~\ref{Fig_smallF}c and \ref{Fig_smallF}d). The first part is given either by $v^*_+$ or $v^*_-$ for $v_0>0$ and $v_0<0$, respectively, under the boundary conditions $(v_0,t_0)$ and $(0,t_{a\pm})$. The third part from $v^*=0$ axis is given by $v^*_+$ or $v^*_-$ for $v_0>$ and $v_0<0$, respectively, under the boundary conditions $(0,t_{b\pm})$ and $(v,t)$.

Indirect paths on one half plane (indicated by the subscript $id$) are thus parametrized by (cf.~Fig.~\ref{Fig_smallF}c)
\be
\label{id_par}
v^*_{id}(s)=
\left\{
\begin{array}{lll} 
\left.v^*_\pm(s)\right|_{v_0,t_0}^{0,t_{a\pm}} & & t_0\le s\le t_{a\pm}\\
0 & & t_{a\pm}<s<t_{b\pm}\\
\left.v^*_\pm(s)\right|_{0,t_{b\pm}}^{v,t} & & t_{b\pm}\le s \le t,
\end{array}\right.
\ee
respectively, while indirect crossing paths (indicated by the subscript $idc$) are parametrized by (cf.~Fig.~\ref{Fig_smallF}d)
\be
\label{idx_par}
v^*_{idc}(s)=
\left\{
\begin{array}{lll} 
\left.v^*_\pm(s)\right|_{v_0,t_0}^{0,t_{a\pm}} & & t_0\le s\le t_{a\pm}\\
0 & & t_{a\pm}<s<t_{b\pm}\\
\left.v^*_\mp(s)\right|_{0,t_{b\pm}}^{v,t} & & t_{b\pm}\le s \le t.
\end{array}\right.
\ee

Clearly, the times $t_{a\pm}$, $t_{b\pm}$ have to satisfy the conditions $t_0< t_{a\pm}\le t_{b\pm}<t$. The actions associated with the indirect paths on one half plane and the indirect crossing paths follow from Eq.~(\ref{id_par}) and Eq.~(\ref{idx_par}) with Eq.~(\ref{lambda})
\be
\label{Aind}
A[\dot{v}^*_{id},v^*_{id}]&=&\Lambda_\pm(0,t_{a\pm};v_0,t_0)+\Lambda_\pm(v,t;0,t_{b\pm})\\
A[\dot{v}^*_{idc},v^*_{idc}]&=&\Lambda_\pm(0,t_{a\pm};v_0,t_0)+\Lambda_\mp(v,t;0,t_{b\pm}),
\ee
where the times $t_{a\pm}$ and $t_{b\pm}$ are then determined by the principle of minimal action. This is a minimization problem subject to the inequality constraint
\be
\label{ineq}
t_0< t_{a\pm}\le t_{b\pm}<t,
\ee
and has two possible solutions depending on the relative positions of the initial and final velocities (cf. \cite{Baule10} for an analogous discussion). These solutions are as follows:

(i) Setting the derivatives of $A[\dot{v}^*_{id},v^*_{id}]$ and $A[\dot{v}^*_{idc},v^*_{idc}]$ with respect to $t_{a\pm}$ and $t_{b\pm}$ equal to zero yields
\be
\label{tbar1}
t_{a\pm}&=&t_0+\tau_m\ln\left(1+\frac{|v_0|}{(\Delta\mp a) \tau_m}\right)\\
\label{tbar2}
t_{b\pm}&=&t-\tau_m\ln\left(1+\frac{|v|}{(\Delta\mp a) \tau_m}\right).
\ee
Note that the times $t_{a\pm}$ are just the times at which a deterministic path, described by Eq.~(\ref{det_eq}), would relax to a zero velocity starting from $(v_0,t_0)$. Likewise, $t_{b\pm}$ are the times at which a deterministic path starting at $(v,t)$ would relax to a zero velocity, moving backward in time. We remark that $t_{a\pm}$ depends only on $v_0$ and not on $v$, while $t_{b\pm}$ depends only on $v$ and not on $v_0$.

The action of the indirect paths can then be calculated explicitly by substituting Eqs.~(\ref{tbar1}) and (\ref{tbar2}) into Eq.~(\ref{Aind}). This yields the simple result
\be
\label{Aind2}
A[\dot{v}^*_{id},v^*_{id}]&=&\Lambda_\pm(0,t_{a\pm};v_0,t_0)+\Lambda_\pm(v,t;0,t_{b\pm})\nonumber\\
&=&\Lambda_\pm(v,t;0,t_{b\pm})\nonumber\\
&=&U(v).
\ee
Likewise,
\be
\label{Aind2b}
A[\dot{v}^*_{idc},v^*_{idc}]&=&\Lambda_\pm(0,t_{a\pm};v_0,t_0)+\Lambda_\mp(v,t;0,t_{b\mp})=U(v).
\ee
The first term, $\Lambda_\pm(0,t_{a\pm};v_0,t_0)$, on the right hand side of the first line of Eq.~(\ref{Aind2}) vanishes, because it is the action associated with the deterministic relaxation of the object and for such paths the Lagrangian and the corresponding actions are identically zero (cf. Eq.~(\ref{lagrangian})). The third line follows upon substituting Eq.~(\ref{tbar2}) into Eq.~(\ref{lambda}) and noting that $\Lambda_+(v,t;0,t_{b+})$ only contributes for positive $v$ and $\Lambda_-(v,t;0,t_{b-})$ only for negative $v$, giving rise to $U(v)$. 

(ii) If the initial and final velocities of the optimal path are such that $t_{b\pm}<t_{a\pm}$ in conflict with Eq.~(\ref{ineq}), the minimum of Eq.~(\ref{Aind}) has to be determined under the equality constraint $t_{a\pm}=t_{b\pm}$. If the initial and final velocities have opposite signs the minimization then yields Eq.~(\ref{tbar_eq}) and the solution is $t^\times$ in this case, i.e., the indirect crossing paths become direct crossing paths.

\begin{figure}
\begin{center}
\begin{tabular}{ll}
\includegraphics[width=7.5cm]{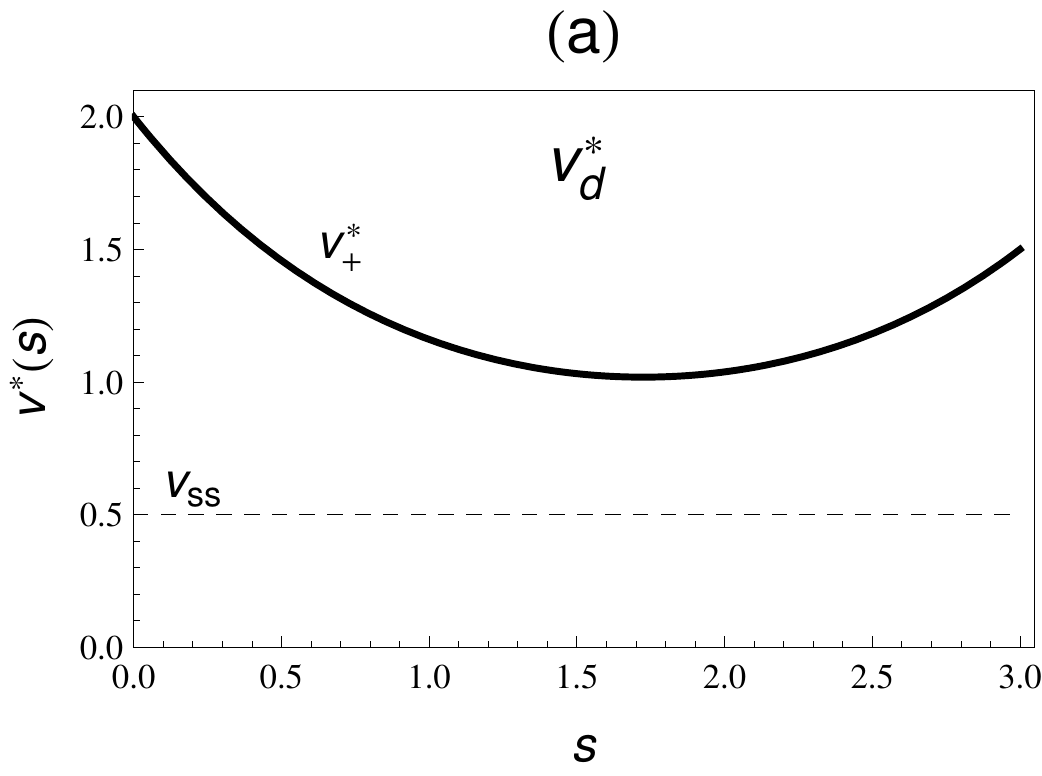} &\hspace{0.5cm}\includegraphics[width=7.5cm]{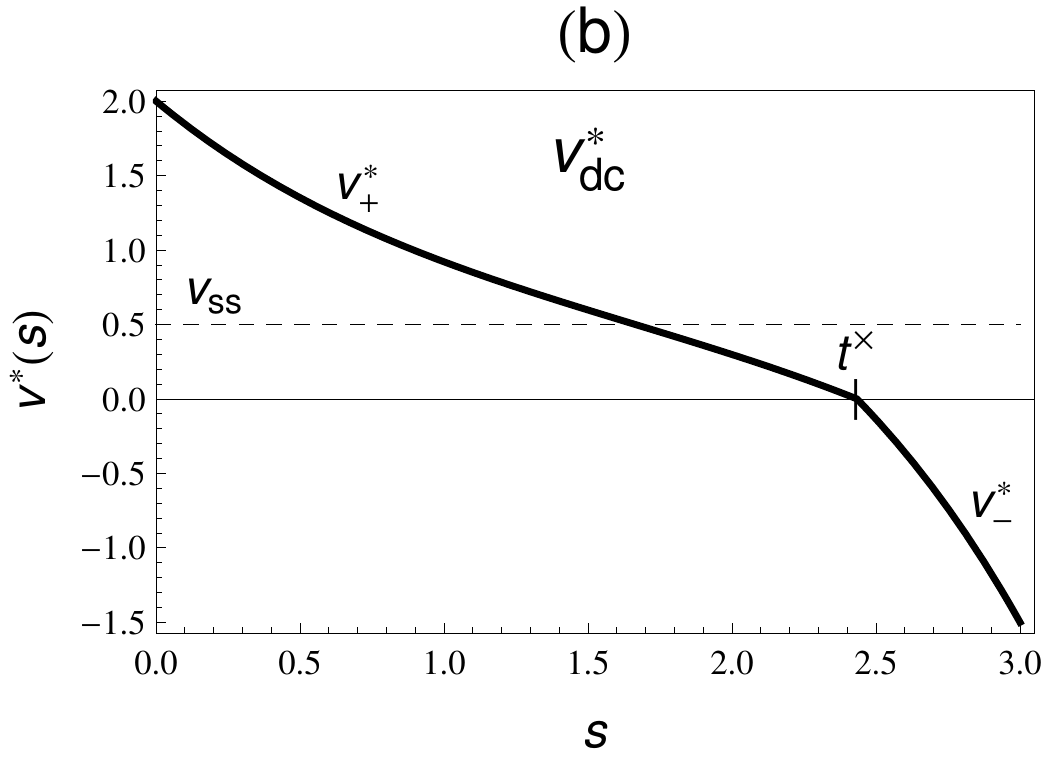}\vspace{0.5cm}\\
\includegraphics[width=7.5cm]{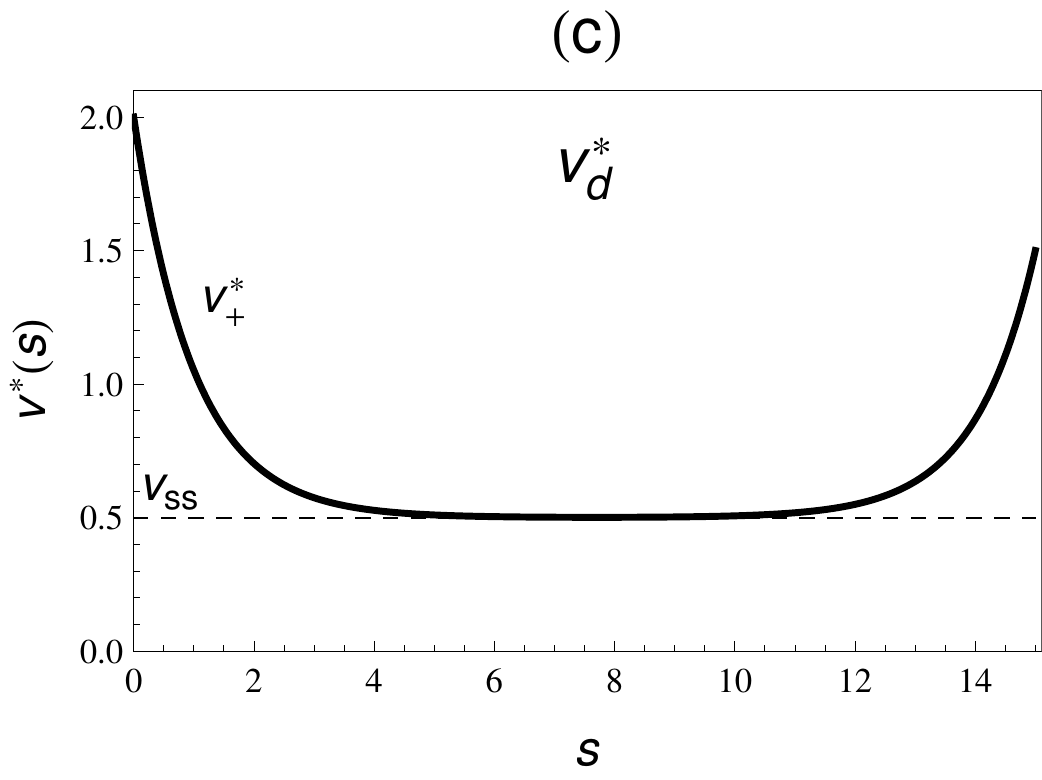} &\hspace{0.5cm}\includegraphics[width=7.5cm]{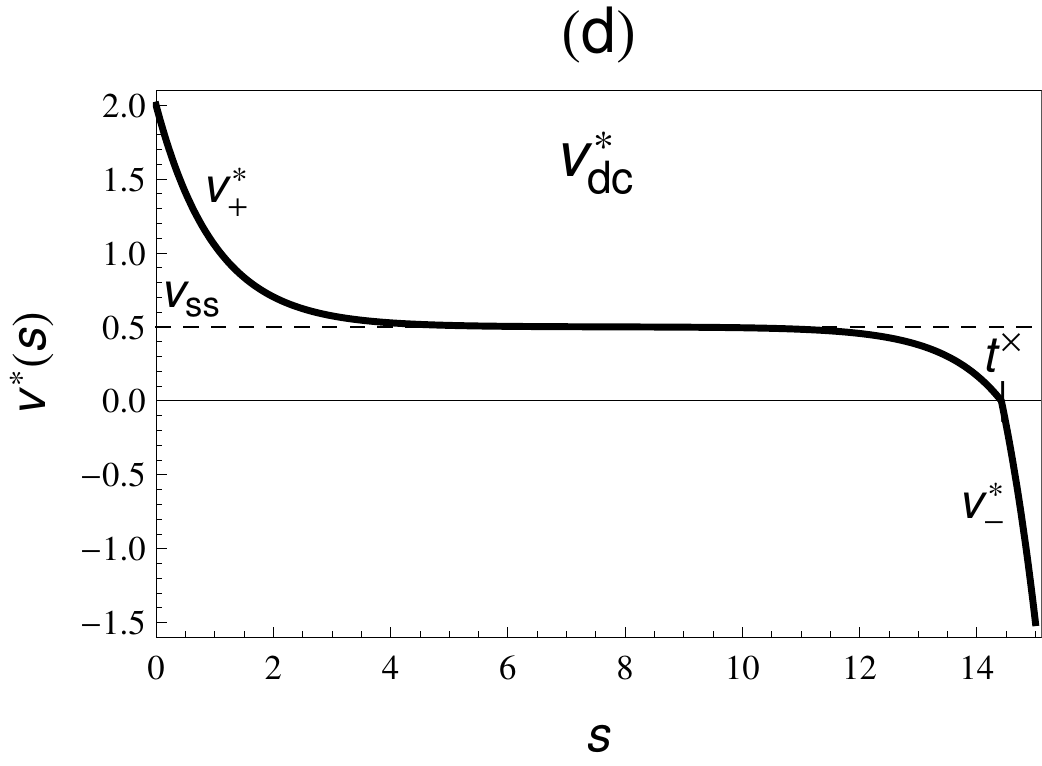}
\end{tabular}
\caption{\label{Fig_largeF}Examples of optimal paths $v^*(s)$ in the case $a>\Delta$ and $v_0>0$. (a) Direct path on one half plane. (b) Direct crossing path. Here, the curvature of $v^*(s)$ changes sign when crossing the $v_{ss}$ line (thin dashed line); cf.~Eq.~(\ref{v_st}). (c), (d) For sufficiently large times, both the direct path on one half plane and the direct crossing path follow the attractor at $v_{ss}$ before reaching their final velocity. Parameter values: $\tau_m=1.0$, $\Delta=0.7$, $a=1.2$.}
\end{center}
\end{figure}

\subsection{The case $a>\Delta$}
\label{ssec_fgd}

\begin{figure}
\begin{center}
\begin{tabular}{ll}
\includegraphics[width=7.5cm]{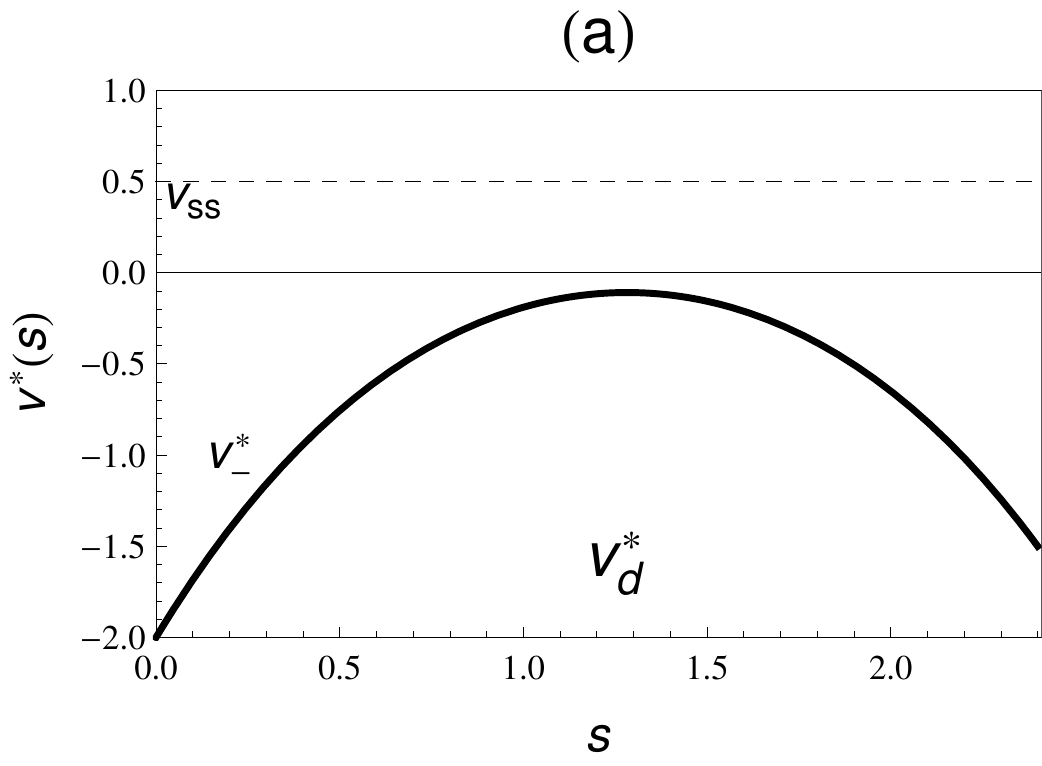} &\hspace{0.5cm} \includegraphics[width=7.5cm]{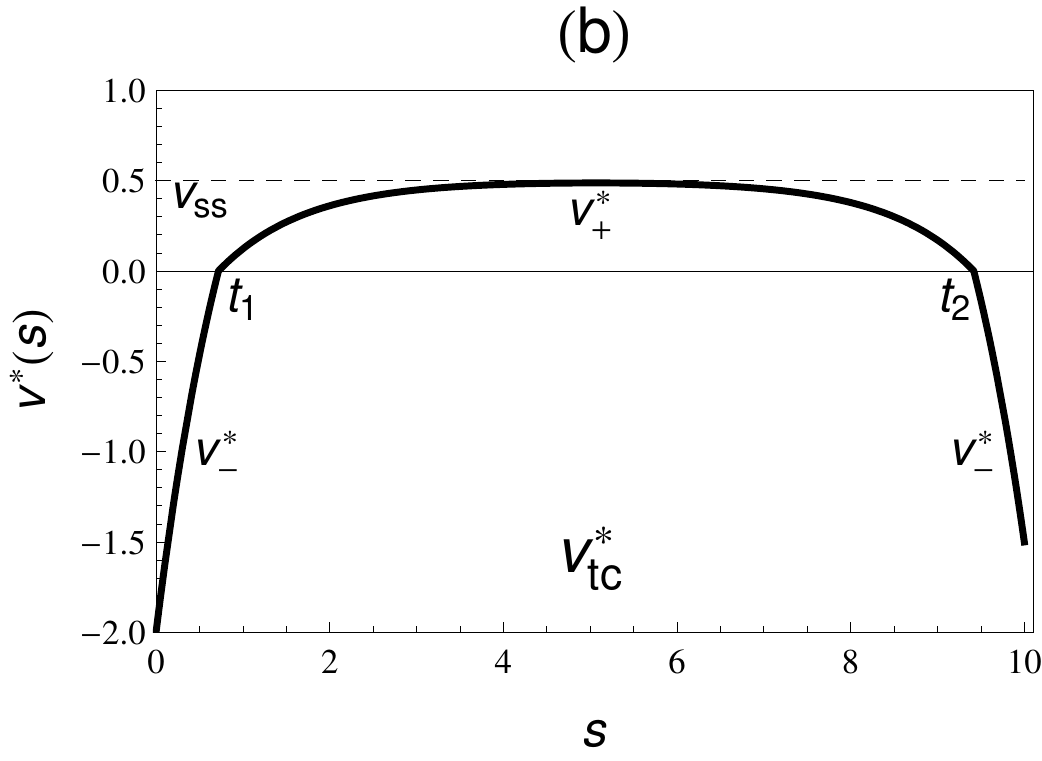}
\end{tabular}
\caption{\label{Fig_dc}Examples of optimal paths in the case $a>\Delta$ if both $v_0<0$ and $v<0$. (a) Direct path on one half plane. (b) For sufficiently large times, the optimal path approaches the attractor before reaching the final velocity. This leads to a path that crosses the $v=0$ axis twice. Parameter values: $\tau_m=1.0$, $\Delta=0.7$, $a=1.2$.}
\end{center}
\end{figure}

In this case, the constant acceleration overcomes permanently the static (dry) friction, so that the object will always exhibit a pure slip motion, while stick-slip motion, represented by indirect paths, does not exist (cf.~Fig.~\ref{Fig_largeF}). The attractor of the dynamics is then given by $v_{ss}$, Eq.~(\ref{v_st}), which is always positive due to our convention $a>0$. If $t-t_0$ is sufficiently larger than the inertial relaxation time $\tau_m$, the optimal path will follow the attractor $v_{ss}$ before reaching the final velocity. One has then to distinguish between three different types of optimal paths depending on the signs of initial and final velocities:

\begin{enumerate}
\item For $v_0>0$ and $v>0$, the optimal path is always given by a direct path on the upper half plane, parametrized by Eq.~(\ref{dir_par}) (cf.~Figs.~\ref{Fig_largeF}a and \ref{Fig_largeF}c).
\item For either $v_0>0$ and $v<0$ or $v_0<0$ and $v>0$, the optimal path is given by a direct crossing path, parametrized by Eq.~(\ref{dirx_par}), where the crossing time $t^\times$ is given as solution of Eq.~(\ref{tbar_eq}) (cf.~Figs.~\ref{Fig_largeF}b and \ref{Fig_largeF}d).
\item For $v_0<0$ and $v<0$, the optimal path can either be a direct path on the lower half plane, parametrized by Eq.~(\ref{dir_par}) (cf.~Fig.~\ref{Fig_dc}a), or it might cross the $v^*=0$ axis \textit{twice} (cf.~Fig.~\ref{Fig_dc}b). The latter case is distinct from the previous cases 1 and 2: first, the optimal path crosses the axis in order to reach the attractor at positive $v_{ss}$, and then goes back to the negative final velocity. The optimal path is thus parametrized by three parts (the subscript $tc$ denotes here ``twice crossing"):
\be
\label{dc_par}
v^*_{tc}(s)=
\left\{
\begin{array}{lll}
\left.v^*_-(s)\right|_{v_0,t_0}^{0,t_1} & & t_0\le s\le t_1\\
\left.v^*_+(s)\right|_{0,t_1}^{0,t_2} & & t_1<s<t_2\\
\left.v^*_-(s)\right|_{0,t_2}^{v,t} & & t_2\le s \le t.
\end{array}\right.
\ee
Here, the two crossing times $t_1$ and $t_2$ are determined by the principle of minimal action. The action of the twice crossing path is clearly:
\be
\label{Adc}
A[\dot{v}^*_{tc},v^*_{tc}]&=&\Lambda_-(0,t_1;v_0,t_0)+\Lambda_+(0,t_2;0,t_1)+\Lambda_-(v,t;0,t_2),
\ee
so that the crossing times are determined by minimization of $A[\dot{v}^*_{tc},v^*_{tc}]$. The crossing times $t_1$ and $t_2$ in this case can only be determined numerically.  
\end{enumerate}

\section{Analytic expressions for the transition probability}
\label{Sec_structure}

In the previous section we have determined the solutions of the E-L equations (\ref{EL}). The transition probability follows from these solutions via the saddlepoint approximation Eq.~(\ref{sp_approx}). However, since different types of optimal paths appear as solutions of the E-L equations (e.g., direct and indirect paths in the case $a<\Delta$), we have to determine from these the one minimizing the action for given initial and final velocities, $(v_0,t_0)$ and $(v,t)$. As before, we discuss the cases $a< \Delta$ and $a>\Delta$ separately. For certain choices of initial and final velocities, it turns out that there are \textit{forbidden} regions in the $(v,s)$-plane which cannot be reached by direct paths but only by indirect paths.

\subsection{The case $a<\Delta$}

In this case, direct paths and indirect paths are possible. We can find the realized optimal path between given initial and final velocities, either direct \textit{or} indirect, as the minimum of their corresponding actions. 

We consider first direct and indirect paths on one half-plane (cf.~Figs.~\ref{Fig_smallF}a,c and Figs.~\ref{Fig_largeF}a,c). The direct paths have the associated action $A[\dot{v}^*_d,v^*_d]$, while the indirect paths have the associated action $A[\dot{v}^*_{id},v^*_{id}]$. From the condition
\be
\label{cond1}
A[\dot{v}^*_{id},v^*_{id}]=A[\dot{v}^*_d,v^*_d],
\ee
which, using Eqs.~(\ref{Adir}) and (\ref{Aind2}), is equivalent to
\be
\label{cond1b}
U(v)=\Lambda_\pm(v,t;v_0,t_0),
\ee
one can derive boundary lines in the $(v,s)$-plane for the final velocity $v(t)$, denoted by $w^+(t)$ and $w^-(t)$,\footnote{The superscripts $+$ and $-$ indicate boundary lines for a positive or a negative initial velocity, respectively.} such that, for $v_0>0$,
\be
A[\dot{v}^*_{id},v^*_{id}] \le A[\dot{v}^*_d,v^*_d] \qquad \mathrm{for} \qquad 0<v\le w^+(t),
\ee
while, for $v_0<0$,
\be
A[\dot{v}^*_{id},v^*_{id}] \le A[\dot{v}^*_d,v^*_d] \qquad \mathrm{for} \qquad w^-(t)\le v<0.
\ee
Equation~(\ref{cond1b}) leads, with the two equations above, to two quadratic equations for $v$, which yield analytic expressions for $w^\pm(t)$:
\be
\label{dpm}
w^\pm(t)&=& \mp\left[(\Delta\mp a)\tau_m-e^{(t-t_0)/\tau_m}((\Delta\mp a)\tau_m+|v_0|)+\sqrt{\left(v_0^2+2(\Delta\mp a)\tau_m |v_0|\right)\left(e^{2(t-t_0)/\tau_m}-1\right)}\right].\nonumber\\
\ee
This implies that if either $0<v\le w^+(t)$ for $v_0>0$ or $w^-(t)\le v<0$ for $v_0<0$, the action of the indirect path is lower than that of the direct path, so that the indirect path will be followed.

Next, we consider direct and indirect crossing paths (cf.~Figs.~\ref{Fig_smallF}b, d and Figs.~\ref{Fig_largeF}b, d). In this case the condition reads
\be
\label{cond2}
A[\dot{v}^*_{idc},v^*_{idc}]=A[\dot{v}^*_{dc},v^*_{dc}],
\ee
and leads to two other critical values or boundary lines for $v(t)$, denoted by $u^+(t)$ and $u^-(t)$, such that for $v_0>0$, i.e., for paths crossing from the upper to the lower half plane,
\be
A[\dot{v}^*_{idc},v^*_{idc}] \le A[\dot{v}^*_{dc},v^*_{dc}] \qquad \mathrm{for} \qquad u^-(t)\le v<0.
\ee
Likewise, for $v_0<0$, i.e., for paths crossing from the lower to the upper half plane, we have the condition
\be
A[\dot{v}^*_{idc},v^*_{idc}] \le A[\dot{v}^*_{dc},v^*_{dc}] \qquad \mathrm{for} \qquad 0<v\le u^+(t).
\ee
Explicit expressions for $u^\pm(t)$ can be obtained without actually solving Eq.~(\ref{cond2}) for $v$. Indeed, the result of the minimization of Eq.~(\ref{Aind}) shows that, when the initial and final velocities are such that $t_{b-}=t_{a+}$, the indirect crossing path becomes a direct crossing path with $t_{b-}=t_{a+}=t^\times$. The condition $t_{b-}=t_{a+}$ leads for $v_0>0$ to a boundary line $u^+(t)$, while for $v_0<0$, the condition $t_{b+}=t_{a-}$ leads to a boundary line $u^-(t)$. Using Eqs.~(\ref{tbar1}) and (\ref{tbar2}) both conditions can be solved for the final velocity $v$, leading to the boundary line
\be
\label{vc_b}
u^\pm(t)=\mp(\Delta\pm a)\tau_m\left(\frac{(\Delta\mp a)\tau_m}{(\Delta\mp a)\tau_m+|v_0|}e^{(t-t_0)/\tau_m}-1\right).
\ee

\begin{figure}
\begin{center}
\begin{tabular}{ll}
\includegraphics[width=7.5cm]{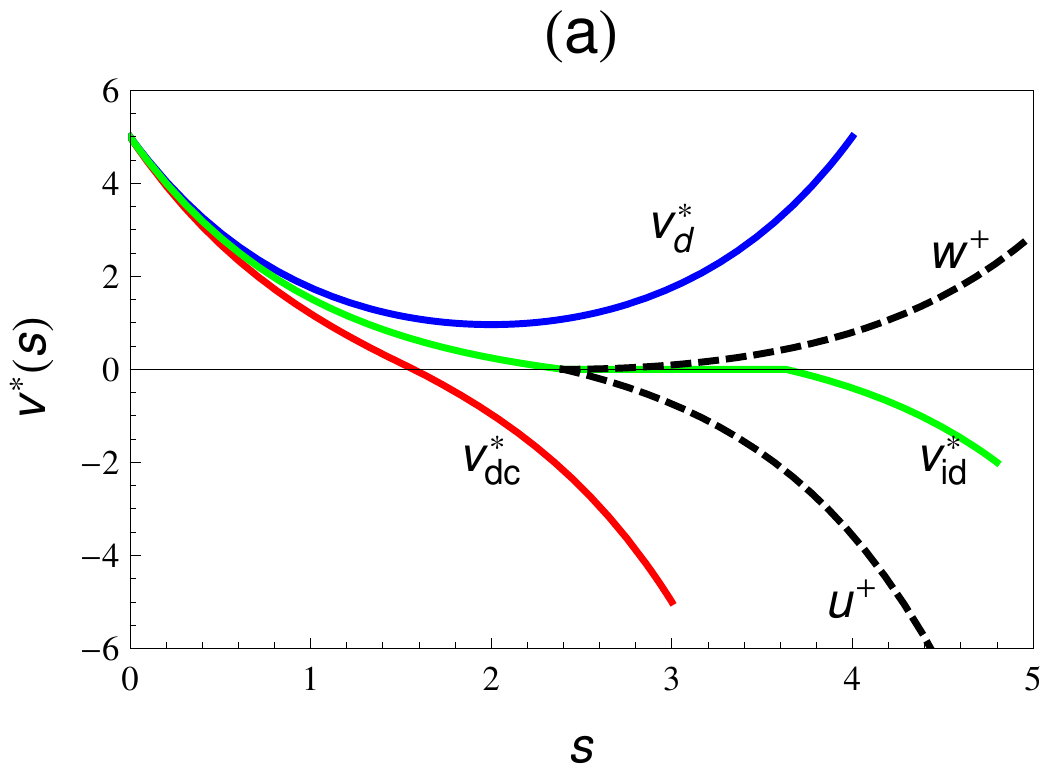} &\hspace{0.5cm} \includegraphics[width=7.5cm]{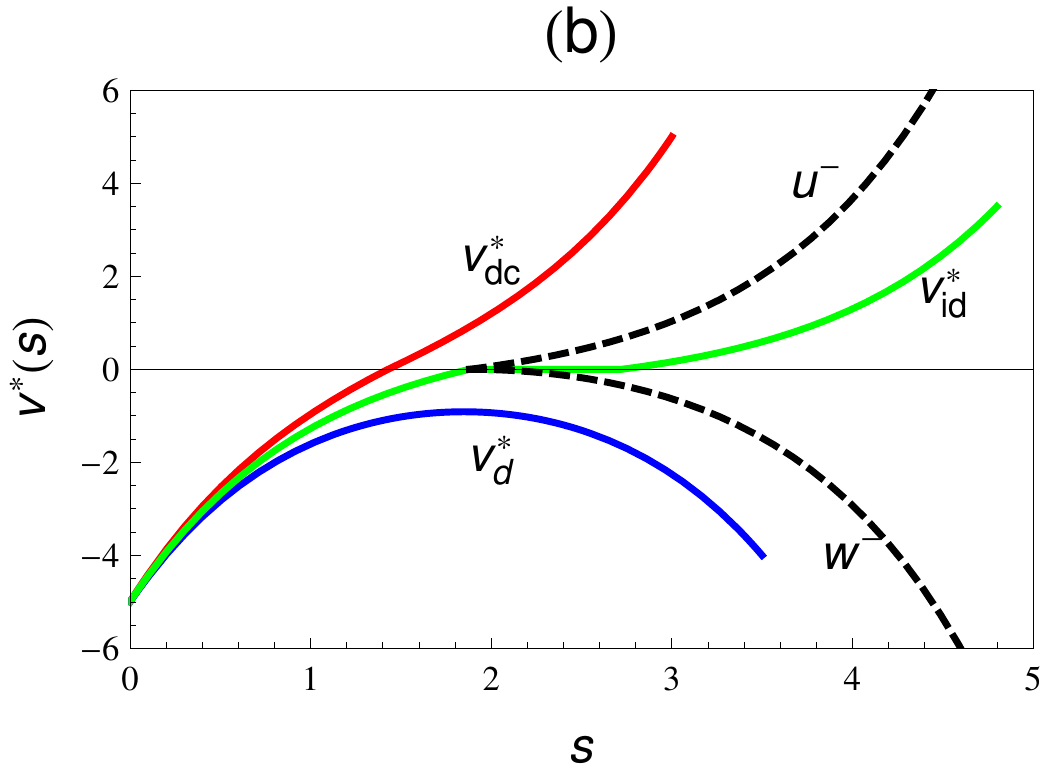}
\end{tabular}
\caption{\label{Fig_bound1}The boundary lines of the optimal paths for the case $a<\Delta$ indicating a change in the dynamical behavior of the object for (a) $v_0>0$ and (b) $v_0<0$. If the final velocity $v$ lies either in the interval $u^+(t)< v<w^+(t)$ ($v_0>0$) or $w^-(t)< v<u^-(t)$ ($v_0<0$), the object follows a stick-slip path since the optimal path is either an indirect path on one half plane or an indirect crossing path (the green curve indicates an indirect crossing path). Otherwise, the optimal path is either a direct or a direct crossing path (blue and red curves, respectively). The boundaries between direct and indirect paths are indicated with dashed black curves. Parameter values: $\tau_m=1.0$, $\Delta=0.7$, $a=0.2$.  }
\end{center}
\end{figure}

It follows from this discussion that for a given initial state $(v_0,t_0)$, the optimal path is an indirect path, either on one half plane or crossing on two half planes, if the final state $(v,t)$ lies in the interval $u^+(t)\le v \le w^+(t)$ ($v_0>0$) or $w^-(t)\le v \le u^-(t)$ ($v_0<0$) with $t\ge t_{a\pm}$, otherwise the optimal path is a direct or a direct crossing path (cf.~Figs.~\ref{Fig_bound1}a and \ref{Fig_bound1}b). The curves $u^\pm(t)$ and $w^\pm(t)$ thus represent boundaries in the $(v,s)$-plane, separating different dynamical behaviors of the moving object in terms of direct and direct crossing paths (pure slip motion) and indirect paths (stick-slip motion).

\begin{figure}
\begin{center}
\begin{tabular}{ll}
\includegraphics[width=7.5cm]{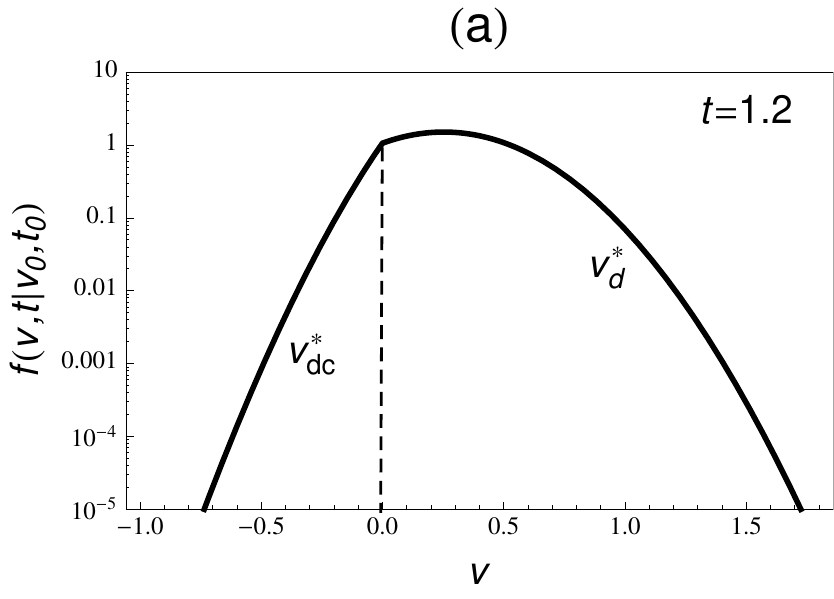} &\hspace{0.5cm}\includegraphics[width=7.5cm]{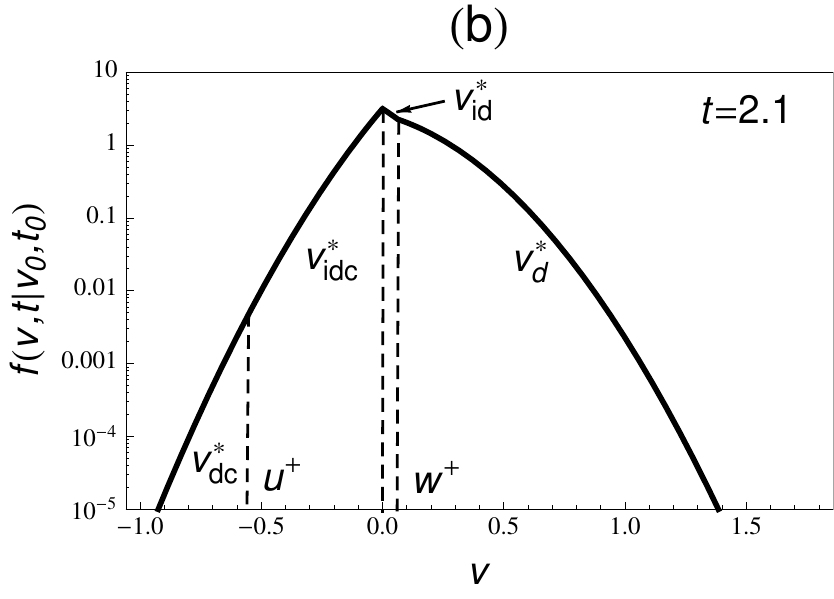}\vspace{0.5cm}\\
\includegraphics[width=7.5cm]{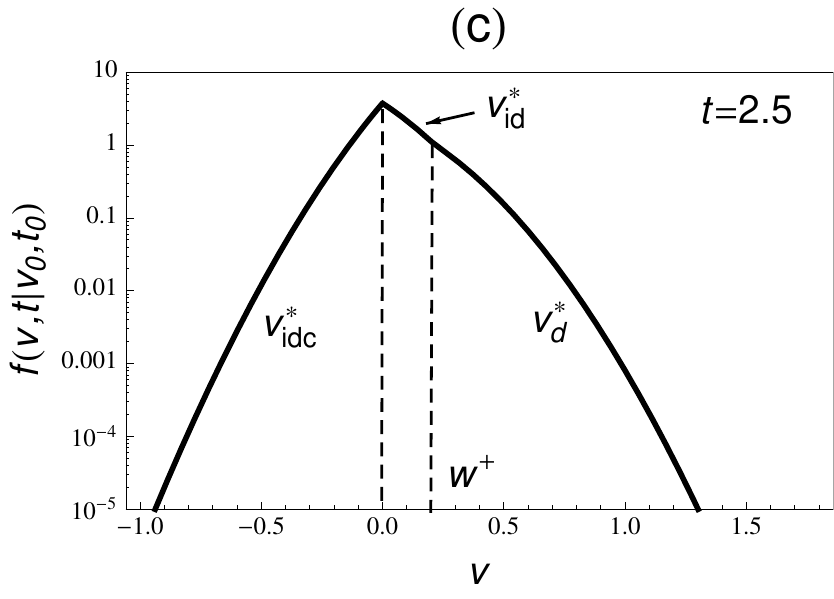} &\hspace{0.5cm}\includegraphics[width=7.5cm]{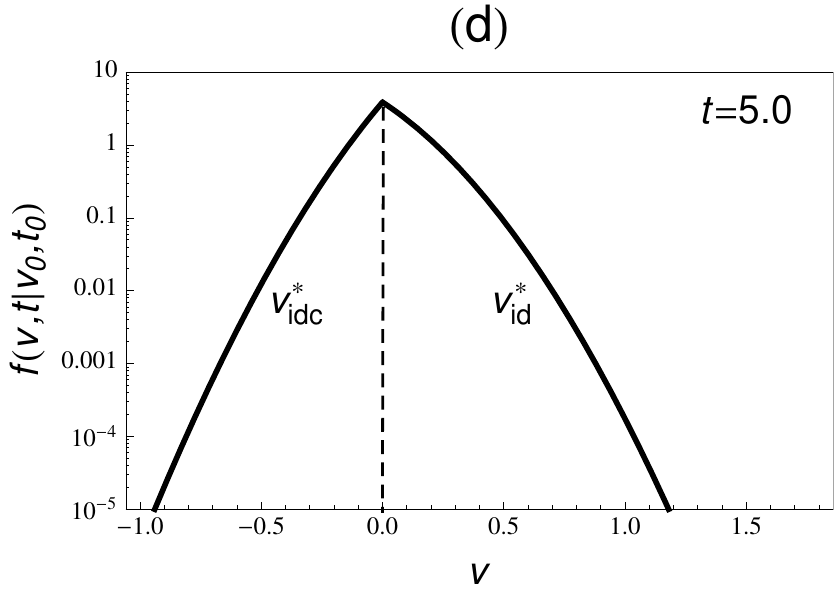}
\end{tabular}
\caption{\label{Fig_dist} Transition probability $f(v,t|v_0,0)$ for $a<\Delta$ and $v_0>0$, as given by Eqs.~(\ref{sol1}) and (\ref{sol2}), plotted for four different times $t$. Transitions between the different types of optimal paths are indicated with vertical dashed lines. At $v=0$, the slope of the transition probability is always discontinuous. (a) For $t<t_{a+}$, only direct crossing and direct paths contribute for $v<0$ and $v>0$, respectively, in which case $f(v,t|v_0,0)$ is given by Eq.~(\ref{sol1}). (b) For $t>t_{a+}$, the distribution is given by Eq.~(\ref{sol2}) and indirect paths then contribute for $u^+<v<w^+$. At the transition between indirect and direct paths, i.e., at $w^+$, the slope of the transition probability is discontinuous. (c) For larger $t$, the indirect paths contribute over a larger region of values for $v$. In this figure the contribution of the direct crossing paths is already far in the tail of $f(v,t|v_0,0)$, so that they do not appear in the figure. (d) For asymptotically large $t$, only indirect paths contribute and the stationary velocity distribution $p_s(v)$, Eq.~(\ref{p_stat}), is recovered. Parameter values: $\tau_m=1.0$, $\Delta=0.7$, $a=0.2$, $\gamma=10.0$, $v_0=2.0$, $t_0=0$, $t_{a+}=1.6$. The values of $u^+$ and $w^+$ can be read off the figures.}
\end{center}
\end{figure}

Having determined the boundaries of the different kinds of optimal paths in the $(v,s)$-plane, we now obtain analytical expressions  for the transition probability $f(v,t|v_0,t_0)$ by using the saddlepoint approximation, Eq.~(\ref{sp_approx}), i.e., by selecting that optimal path, which minimizes the action. For simplicity, we give the results for $v_0>0$ only. The results for $v_0<0$ follow similarly. We distinguish two cases:

(a) For $t\le t_{a+}$ no indirect optimal paths can occur, so that, for a final velocity $v\le 0$, the direct crossing path $v^*_{dc}$, with an action given by Eq.~(\ref{Adirx}), is the optimal path. For $v>0$, on the other hand, the direct path on the upper half plane, with an action given by Eq.~(\ref{Adir}), is the optimal path. By combining these two cases, we thus obtain for the transition probability
\be
\label{sol1}
f(v,t|v_0,t_0)=\mathcal{N}_1
\left\{
\begin{array}{lll}
e^{-\gamma \left[\Lambda_+(0,t^\times;v_0,t_0)+\Lambda_-(v,t;0,t^\times)\right]} & & v\le 0\\
e^{-\gamma \Lambda_+(v,t;v_0,t_0)} & & v>0,
\end{array}\right.
\ee
in the case $t\le t_{a+}$, where $\mathcal{N}_1$ is a normalization factor (cf.~Fig.~\ref{Fig_dist}a).

(b) For $t>t_{a+}$, indirect paths occur and the boundary lines of the optimal paths indicate that
\begin{enumerate}
\item For a final velocity $v\le u^+(t)$, the direct crossing paths with action given by Eq.~(\ref{Adirx}) are optimal;

\item For $u^+(t)<v<0$, the indirect crossing with action given by  Eq.~(\ref{Aind2b}) are optimal;

\item For $0<v<w^+(t)$, the indirect paths on the upper half plane are optimal. The actions of these paths are given by Eq.~(\ref{Aind2}) and are the same as those of indirect crossing paths;

\item For $v\ge w^+(t)$, the direct paths with action given by Eq.~(\ref{Adir}) are optimal.
\end{enumerate}
Combining these cases, we then obtain
\be
\label{sol2}
f(v,t|v_0,t_0)=\mathcal{N}_2
\left\{
\begin{array}{lll}
e^{-\gamma \left[\Lambda_+(0,t^\times;v_0,t_0)+\Lambda_-(v,t;0,t^\times)\right]} & & v\le u^+\\
e^{-\gamma U(v)} & & u^+<v<w^+\\
e^{-\gamma  \Lambda_+(v,t;v_0,t_0) } & & v\ge w^+,
\end{array}\right.
\ee
where $\mathcal{N}_2$ is a normalization factor (cf.~Fig.~\ref{Fig_dist}b-d). 

Note that the contribution of the indirect paths for $u^+<v<w^+$ is just the stationary distribution Eq.~(\ref{p_stat}). This implies that we recover the stationary distribution in the asymptotic time limit from Eq.~(\ref{sol2}),
\be
\lim_{t\rightarrow\infty}f(v,t|v_0,t_0)=p_s(v),
\ee
since then both $w^+\rightarrow \infty$ and $u^+\rightarrow-\infty$ as $t\rightarrow\infty$, respectively, so that only indirect paths contribute to the transition probability in the asymptotic time limit.

The transition probability $f(v,t|v_0,t_0)$ is plotted in Fig.~\ref{Fig_dist} for four different $t$ values. One of the dominant features of the transition probability is the discontinuity in the slope at $v=0$, which appears for all $t$ values and is a direct consequence of the $\sigma(v)$ singularity in the equation of motion (\ref{bm}). In addition, the slope of the transition probability is also discontinuous at the boundary line $w^+$, where the optimal path switches from an indirect to a direct path, i.e., to a different dynamical behavior.

\subsection{The case $a>\Delta$}

It follows from the discussion of the optimal paths given in Sec.~\ref{ssec_fgd} that, for $v_0>0$, the optimal path is a direct path if the final velocity $v$ is also positive (case 1 in Sec.~\ref{ssec_fgd}), whereas the optimal path is a direct crossing path if $v<0$ (case 2 in Sec.~\ref{ssec_fgd}). Therefore, for $v_0>0$, the transition probability is given by Eq.~(\ref{sol1}) \textit{for all $t$}.

The situation is different for $v_0<0$. If the final velocity $v$ is positive, then the optimal path will just be a direct crossing path (case 2 in Sec.~\ref{ssec_fgd}). However, if $v<0$, two types of paths are possible, namely a direct and a \textit{twice} crossing path (case 3 in Sec.~\ref{ssec_fgd}). Which one is chosen depends on the location of the final velocity and is determined by the principle of minimal action. From the condition
\be
\label{cond3}
A[\dot{v}^*_{d},v^*_{d}]=A[\dot{v}^*_{tc},v^*_{tc}],
\ee
follows then a \textit{critical} value of $v$, denoted by $w_{cr}(t)$, such that
\be
A[\dot{v}^*_{d},v^*_{d}]\le A[\dot{v}^*_{tc},v^*_{tc}] \qquad \mathrm{for} \qquad w_{cr}(t)\le v<0.
\ee
This indicates that
\begin{enumerate}
\item For a final velocity $v< w_{cr}(t)$, the direct paths with action (\ref{Adir}) are optimal;
\item For $w_{cr}(t)<v<0$, the twice crossing ($tc$) paths with action (\ref{Adc}) are optimal;
\item For $v\ge 0$, the direct crossing paths with action  (\ref{Adirx}) are optimal.
\end{enumerate}

From these results and Eq.~(\ref{sp_approx}), we thus obtain the transition probability for $v_0<0$:
\be
\label{sol3}
f(v,t|v_0,t_0)=\mathcal{N}_3
\left\{
\begin{array}{lll}
e^{-\gamma \Lambda_-(v,t;v_0,t_0)} & & v\le w_{cr}\\
e^{-\gamma \left[\Lambda_-(0,t_1;v_0,t_0)+\Lambda_+(0,t_2;0,t_1)+\Lambda_-(v,t;0,t_2)\right]} & & w_{cr}<v<0\\
e^{-\gamma \left[\Lambda_-(0,t^\times;v_0,t_0)+\Lambda_+(v,t;0,t^\times)\right]  } & & v\ge 0,
\end{array}\right.
\ee
where $\mathcal{N}_3$ is a normalization factor.

\section{The stationary distribution of the mechanical work}

In this section, we investigate the properties of the mechanical work, performed on the object of mass $m$ due to the external force $F=m\,a$ in the NESS. This work is given as a functional of the object's velocity, Eq.~(\ref{work}), and is therefore a random variable. In the following, we calculate the probability distribution of the mechanical work $W_\tau$ performed over a time $\tau>0$ by the external force. Since this quantity is expected to scale proportionally to $\tau$, we actually consider, from now on, the work per unit time or work rate
\be
w=W_\tau/\tau.
\label{work_power}
\ee
The probability distribution associated with this quantity is denoted by $P_\tau(w)$, and is calculated in the asymptotic time limit $\tau\ra\infty$. From this calculation, we also derive an important result characterizing the NESS, referred to as a fluctuation relation.

\subsection{Distribution of the mechanical work}

The path integral representation of $P_\tau(w)$ is
\be
P_\tau(w)=\int J[v(s)]\, \E^{-\gamma A[\dot v(s),v(s)]}\, \delta\left(W_\tau[v(s)]-\tau w\right)\mathcal{D}v(s).
\label{eqpi1}
\ee
Similarly as we did for the propagator $f(v,t|v_0,t_0)$, we shall approximate the path integral of $P_\tau(w)$ by its dominant path, which corresponds here to the path that minimizes the action $A$ subject to the constraint that the work rate along this path has the value $w$. The expression of $P_\tau(w)$ resulting from this approximation is expected to be accurate in the regime where both $\tau\ra\infty$ and $\gamma\ra\infty$, i.e., where the integration time $\tau$ is long and the noise power $D$ of the external vibrations is small. To be more precise, we expect, following the theory of large deviations \cite{Freidlin84} (see also \cite{Touchette09b} and references therein), to obtain an approximation for $P_\tau(w)$ of the form
\be
\label{largeDev}
P_\tau(w)\approx \E^{-\tau\gamma I(w)},\qquad \tau\ra\infty, \gamma\ra \infty,
\ee  
where $I(w)$ is a function, called the \emph{rate function}, which does not depend on $\gamma$ or $\tau$. Given that this approximation is consistent with taking the optimal path of the full path integral of Eq.~(\ref{eqpi1}), $I(w)$ must be given by
\be
I(w)=\lim_{\tau\ra\infty}\frac{1}{\tau} \min A[\dot v(s),v(s)],
\label{eqi1}
\ee
where the minimum of the action, given by Eqs.~(\ref{action}) and (\ref{lagrangian}), has to be determined among all paths $v(s)$ over the time interval $[0,\tau]$ which give rise to the work value $W_\tau[v(s)]=w\tau$. No limit involving $\gamma$ appears in the above expression because $\gamma$ is already factored out of the action; see Eq.~(\ref{eqpi1}). Equivalently, one can express the rate function by
\be
\label{rate_func}
I(w)=\lim_{\tau\ra\infty} \frac{A[\dot v^*_w,v^*_w]}{\tau},
\ee 
if we denote by $v^*_w(s)$ the optimal path that minimizes $A$ subject to the constraint $W_\tau[v^*_w]=w\tau$. As we did for the propagator, the correction that comes from the Jacobian in the path integral is neglected as $\gamma\ra\infty$. 

To solve the constrained minimization problem of Eq.~(\ref{eqi1}), we transform it, following Taniguchi and Cohen \cite{Taniguchi07}, to an unconstrained minimization problem for the modified action 
\be
\label{Keq}
K_\tau[\dot v(s),v(s)]=A[\dot v(s),v(s)]-\beta W_\tau [v(s)],
\ee
which involves a Lagrange multiplier $\beta$ associated with the constraint $W_\tau[v(s)]=w\tau$. Using Eqs.~(\ref{action}) and (\ref{work}), the modified action Eq.~(\ref{Keq}) can be expressed in terms of a modified Lagrangian
\be
\label{modLag}
\mK(\dot v(s),v(s))=\mL(\dot v(s),v(s))-\beta ma v(s),
\ee
so that\footnote{We set $t_0=0$, without loss of generality.}
\be
K_\tau[\dot v(s),v(s)]=\int_{0}^{\tau} \mK(\dot v(s),v(s))\D s.
\ee 
The paths minimizing the modified action $K_\tau$ are then found by solving the functional equation $\delta K_\tau=0$ or, equivalently, by solving the E-L equation
\be
\frac{\D}{\D s}\frac{\partial\mK}{\partial \dot v^*_\beta}-\frac{\partial\mK}{\partial v^*_\beta}=0
\label{eqelk1}
\ee
for the optimal path $v^*_\beta(s)$, where the subscript $\beta$ indicates the dependence of the optimal path on the Lagrange multiplier $\beta$. This E-L equation must be solved over a time interval $[0,\tau]$ with a given initial velocity $v^*_\beta(0)=v_0$ and a final velocity $v^*_\beta(\tau)=v$. The value of $\beta$ is found by solving the constraint equation $W_\tau[v^*_\beta]=w\tau$ for $\beta$. The Lagrange multiplier $\beta$ is a function of the work rate $w$ only, i.e., $\beta=\beta(w)$, and the optimal path $v^*_w$, required to determine the rate function in Eq.~(\ref{rate_func}), is given by $v^*_w=v^*_{\beta(w)}$.

\begin{figure}
\begin{center}
\begin{tabular}{ll}
\includegraphics[width=7.5cm]{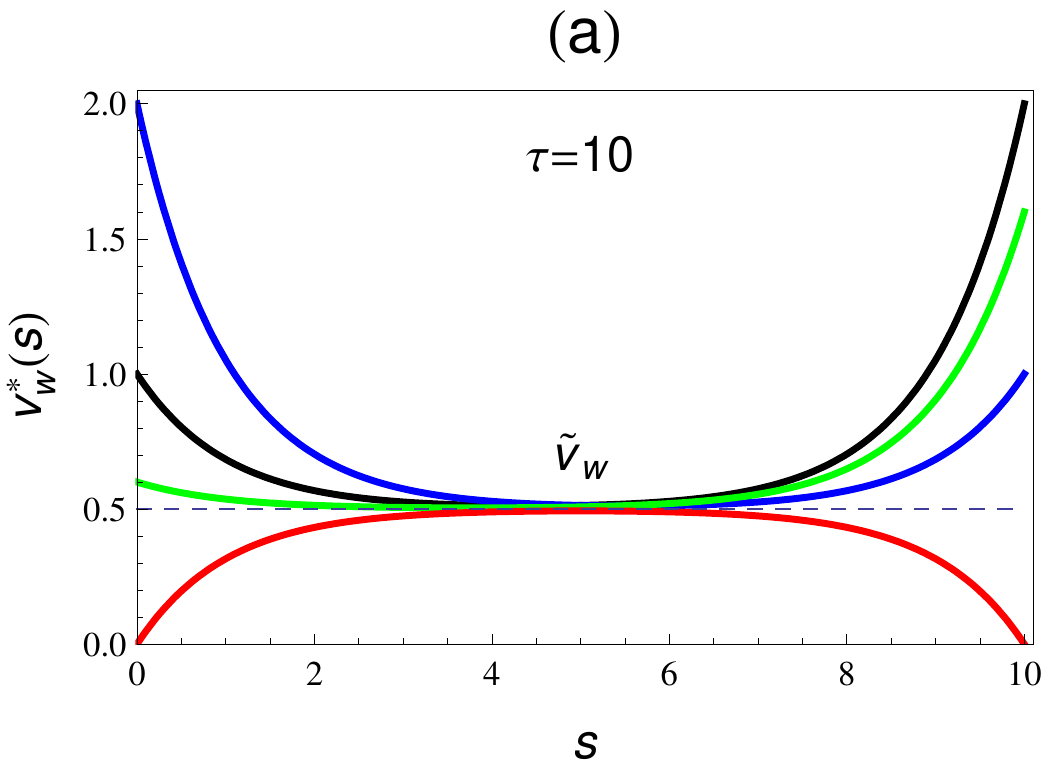} &\hspace{0.5cm}\includegraphics[width=7.5cm]{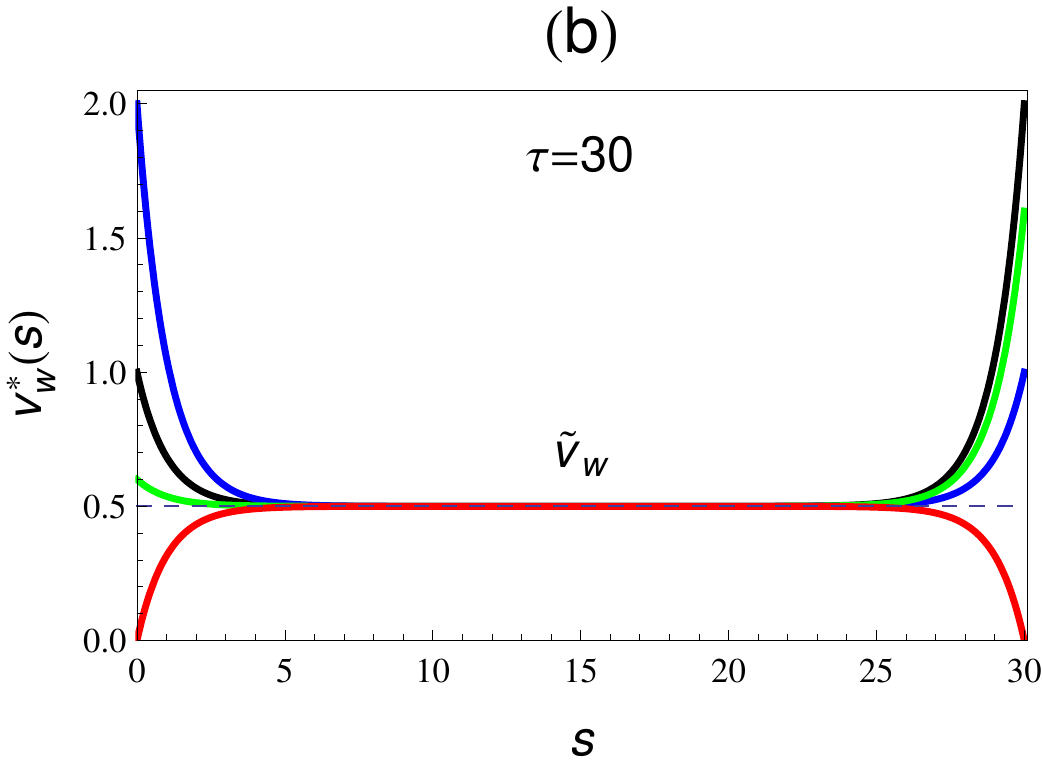}\vspace{0.5cm}\\
\includegraphics[width=7.5cm]{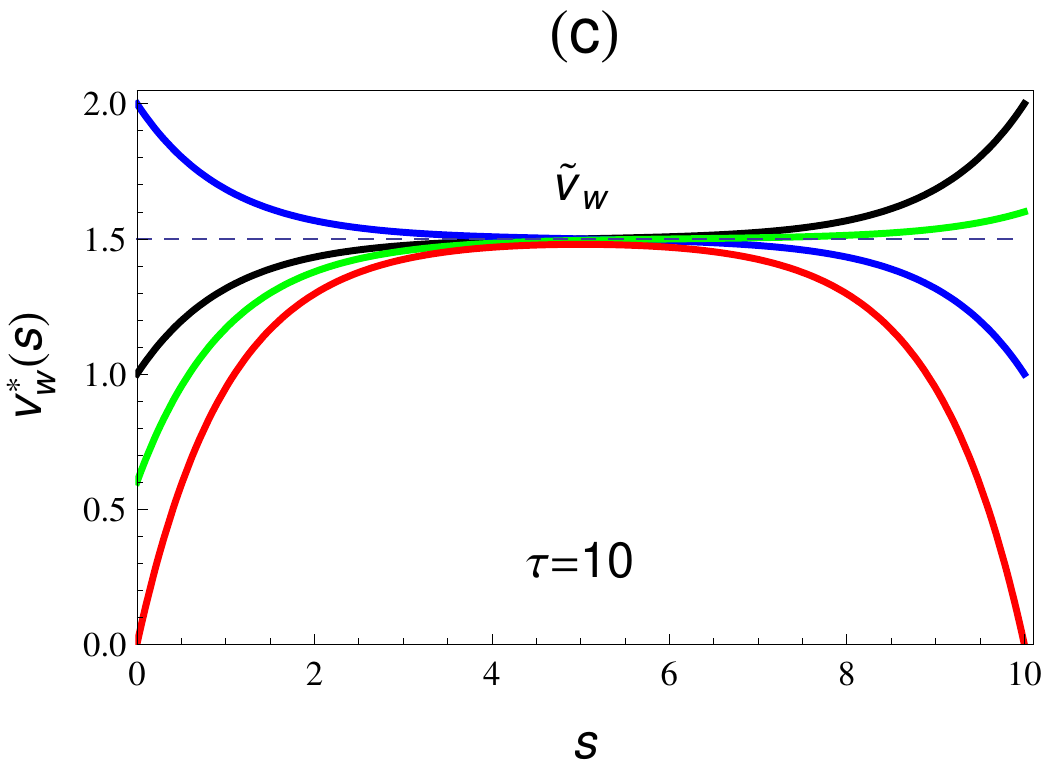} &\hspace{0.5cm}\includegraphics[width=7.5cm]{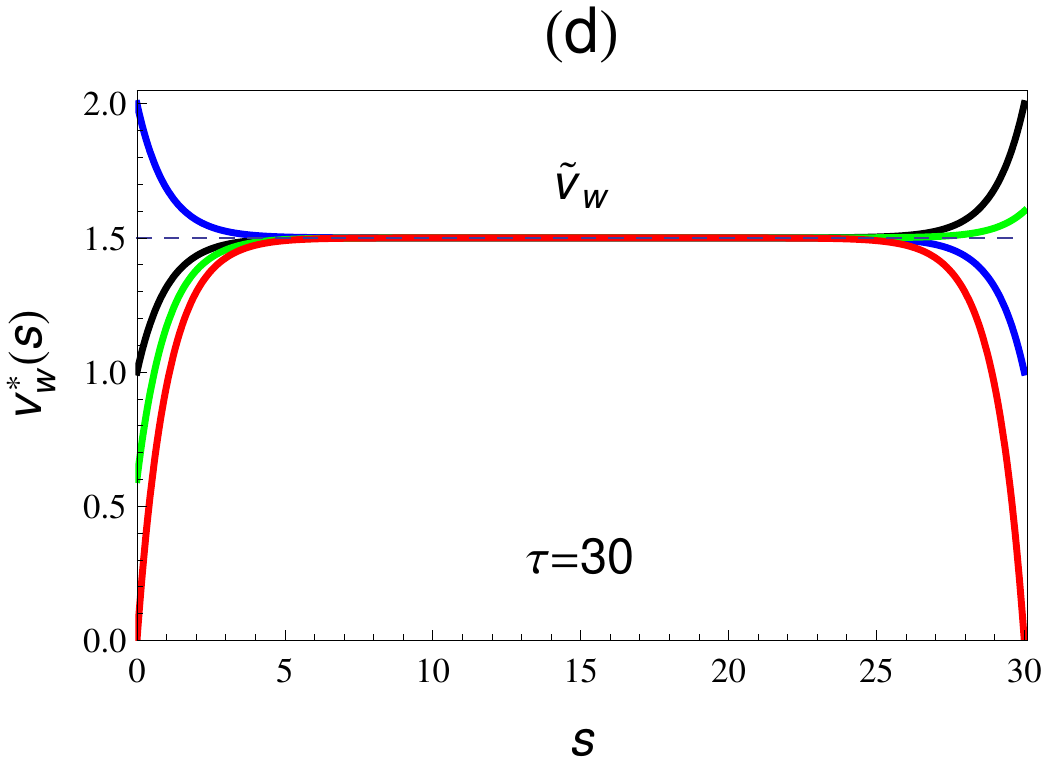}
\end{tabular}
\caption{\label{Fig_workpath}Examples of optimal paths $v^*_w(s)$ that minimize the action $K_\tau$, Eq.~(\ref{Keq}). The different paths are distinguished by different initial and final velocities, which can be read off the figure, e.g., the blue curve is the optimal path with the initial velocity $v_0=2.0$ and the final velocity $v=1.0$ at the time $\tau=10$, etc. For sufficiently large $\tau$, the optimal path is close to a constant velocity $\tilde{v}_w$ (dashed horizontal line), which is the attractor of the constrained motion described by the modified Lagrangian Eq.~(\ref{modLag}). (a) $w=2.0$ and $\tau=10$. (b) $w=2.0$ and $\tau=30$. The time spent close to the attractor $\tilde{v}_w$ scales with $\tau$, while the times that the optimal path needs to reach $\tilde{v}_w$ from the initial velocity or to reach the final velocity from $\tilde{v}_w$, respectively, do not. (c) $w=4.0$ and $\tau=10$. (d) $w=4.0$ and $\tau=30$. Parameter values: $\tau_m=m=\Delta=1.0$, $a=0.5$.}
\end{center}
\end{figure}

Figure~\ref{Fig_workpath} shows typical optimal paths $v^*_w$ that minimize the action $K_\tau$ for different initial and final conditions as well as for different values of $w$. The essential property of these paths, which is clearly seen in this figure, is that they are attracted to a constant value $\tilde{v}_w$, and deviate from $\tilde{v}_w$ only near the initial and final velocities. This behavior of $v^*_w$ follows because the Euler-Lagrange dynamics determined by Eq.~(\ref{eqelk1}) has a unique attractor at $\tilde{v}_w$, whose value depends on $w$. As a result, for sufficiently large $\tau$, all optimal paths satisfying the E-L equation~(\ref{eqelk1}) converge to the attractor $\tilde{v}_w$ starting from any initial velocity $v_0$, then remain near $\tilde{v}_w$ for most of the time, and depart from $\tilde{v}_w$ near the end of the time interval $[0,\tau]$, to join the final velocity $v(\tau)=v$. Figure~\ref{Fig_workpath} also shows that the time that the optimal path needs to reach $\tilde{v}_w$ from $v_0$ as well as to reach $v$ from $\tilde{v}_w$, respectively, do not scale with $\tau$. This implies that the time the optimal path spends near the attractor $\tilde{v}_w$ scales with $\tau$. 

The conclusion that we reach from these observations is that, as $\tau\ra\infty$, the main contribution to the action $A[\dot v^*_w,v^*_w]$ comes increasingly from that part of $v^*_w$ which follows the attractor $\tilde{v}_w$. For the purpose of calculating the rate function $I(w)$, which requires the limit $\tau\ra\infty$ (cf. Eq.~(\ref{rate_func})), we can therefore assume that the optimal path has the simple form
\be
\label{vw}
v^*_w(s)=\tilde{v}_w,
\ee
i.e., is constant for all $s$, but with the constant depending on the work rate fluctuation $w$. 

\begin{figure}
\begin{center}
\begin{tabular}{ll}
\includegraphics[width=7.5cm]{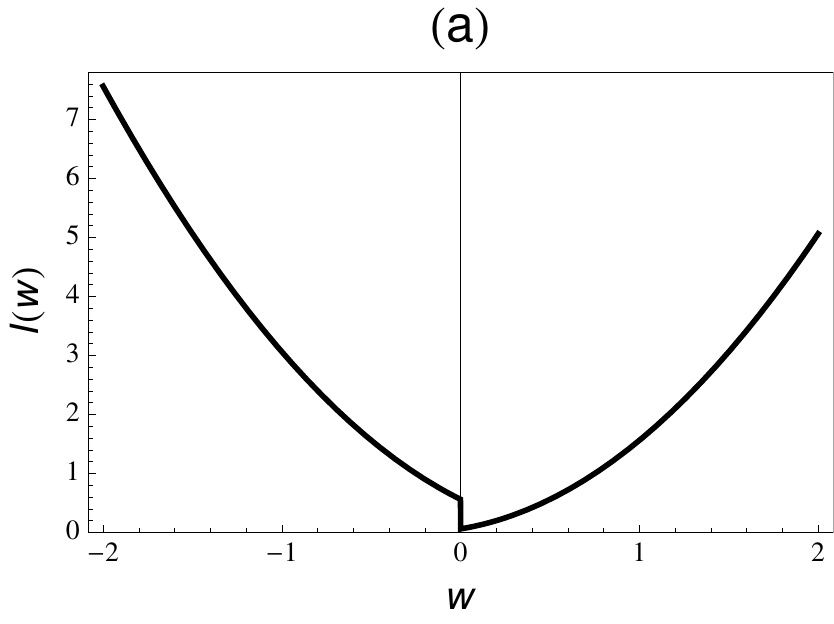} &\hspace{0.5cm}\includegraphics[width=7.9cm]{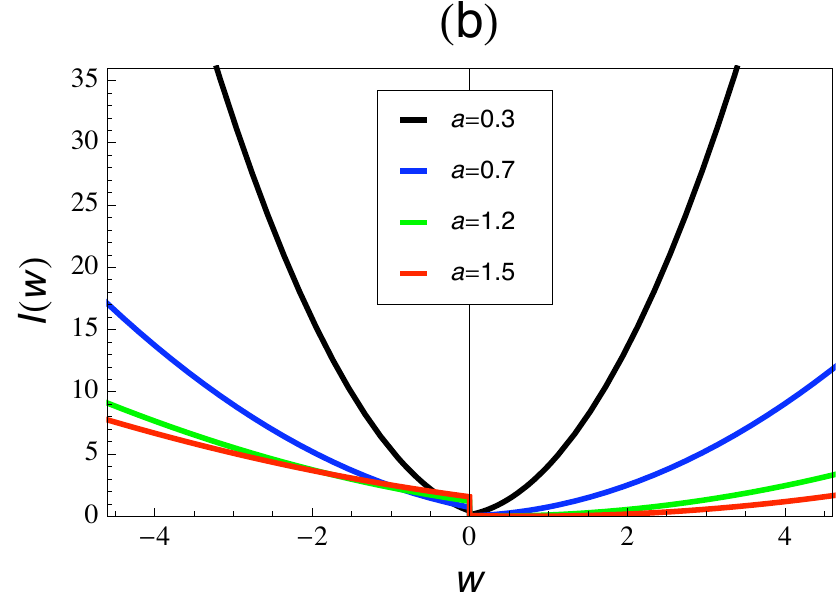}
\end{tabular}
\caption{\label{Fig_ratef}The rate function $I(w)$ of Eq.~(\ref{Iw}) with a discontinuity at $w=0$. The jump at $w=0$ has the magnitude $a\Delta$. Parameter values: $\tau_m=1.0$, $\Delta=1.0$. (a) For $a=0.5$. (b) For various $a$ values.}
\end{center}
\end{figure}

To find the actual value of $\tilde{v}_w$ associated with a given work fluctuation $w$, we substitute Eq.~(\ref{vw}) into the definition of the work, Eq.~(\ref{work}), to obtain
\be
W_\tau[\tilde{v}_w]=\tau ma \tilde{v}_w.
\ee
Using Eq.~(\ref{work_power}) therefore leads to the simple result
\be
\label{wF}
\tilde{v}_w=\frac{w}{ma}.
\ee
The rate function $I(w)$ is then obtained from Eq.~(\ref{rate_func}) by substituting the constant optimal path of Eq.~(\ref{vw}) with Eq.~(\ref{wF}) in the expression of the action, which leads to a constant Lagrangian term:
\be
\label{Iw}
I(w)=\lim_{\tau\ra\infty}\frac{1}{\tau} \int_0^\tau \mL(\dot{\tilde{v}}_w,\tilde{v}_w)\, \D s=\mL(0,\tilde{v}_w)=\frac{1}{4}\left(\frac{w}{ma\tau_m }+\sigma\left(\frac{w}{ma}\right)\Delta-a\right)^2.
\ee
This rate function is plotted in Fig.~\ref{Fig_ratef}. We can see that $I(w)$ has a discontinuity at $w=0$, which is due to the singular nature of the dry friction at $v=0$. The jump of $I(w)$ at $w=0$ is equal to $a\Delta$. 

At this point, it is unclear whether the jump of $I(w)$ at $w=0$ is a real feature of $P(W_\tau)$ or only an artifact of the large deviation approximation. On physical grounds, we expect this distribution to be continuous at $w=0$, and it may well be, accordingly, that its large deviation approximation must be constructed, as for optimal paths, by considering the regions $v>0$ and $v<0$ separately, and by matching at the origin the results obtained. Settling this issue would involve a study of the contribution of the Jacobian correction term, Eq.~(\ref{jacobian}), in the large deviation approximation (\ref{largeDev}), which is beyond the scope of this paper.

\subsection{Fluctuation Relation for the mechanical work}

Fluctuation relations are mathematical relations for the fluctuations of thermodynamic quantities, such as heat (entropy production) or work in nonequilibrium systems (see, e.g., \cite{Harris07} and references therein). In a NESS, the so called asymptotic or steady state fluctuation theorem (SSFT) states that the probability distribution $P_\tau(A)$ of finding a particular value of a thermodynamic quantity $A$ over a time $\tau$ satisfies a certain symmetry relation of the form (cf. \cite{EvansD93,Gallavotti95,Kurchan98,Lebowitz99})
\be
\label{conventional}
\frac{P_\tau(A)}{P_\tau(-A)}\approx e^{c\tau A},
\ee
where $c$ is a constant. The SSFT Eq.~(\ref{conventional}), also refered to as conventional or Gallavotti-Cohen fluctuation theorem \cite{Gallavotti95}, represents a refinement of the second law of thermodynamics in that it quantifies the probability of observing fluctuations of thermodynamic quantities around their average values in a NESS.

\begin{figure}
\begin{center}
\includegraphics[width=7.5cm]{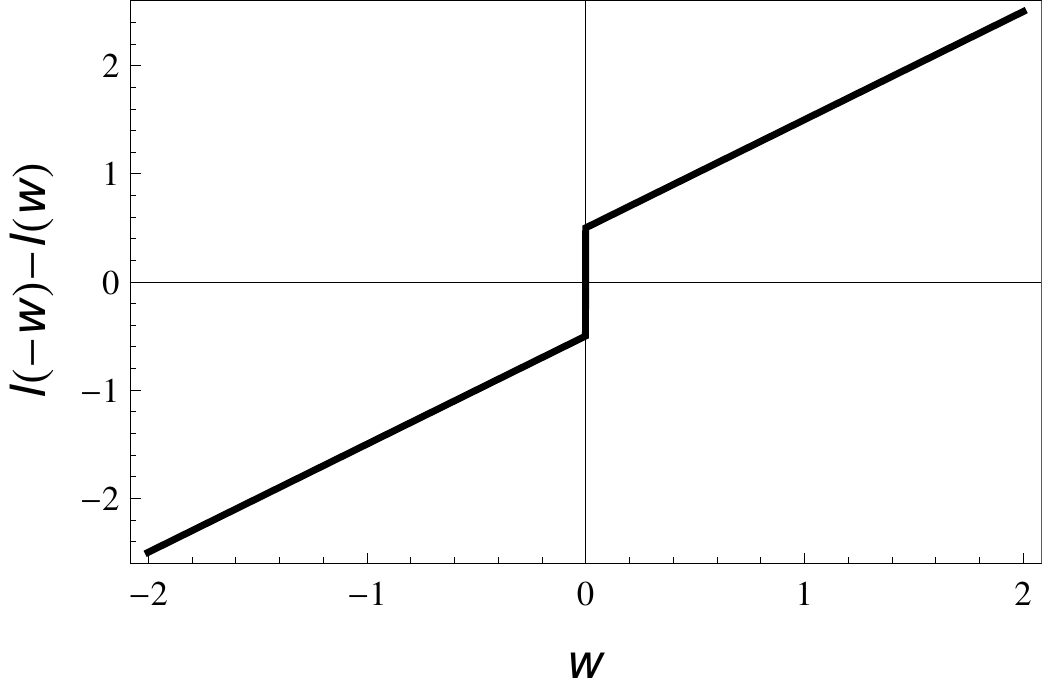}
\caption{\label{Fig_dryft}The difference $I(-w)-I(w)$, Eq.~(\ref{ftexp}). The jump at $w=0$ has the magnitude $2a\Delta$, i.e., twice the jump of $I(w)$ at $w=0$. Parameter values: $\tau_m=m=1.0$, $\Delta=1.0$, $a=0.5$. }
\end{center}
\end{figure}

Using our analytical result for the rate function $I(w)$, Eq.~(\ref{Iw}), we can derive a fluctuation relation for the fluctuations of the mechanical work due to the external force in the stationary regime.\footnote{Importantly, these work fluctuations are \textit{macroscopic} fluctuations, induced by the mechanical vibrations of the plate due to the externally imposed noise.} The ratio of positive to negative work fluctuations in this case is just given by the difference of the corresponding rate functions (cf. Eq.~(\ref{largeDev}))
\be
\frac{P_\tau(w)}{P_\tau(-w)}\approx e^{\tau\gamma[I(-w)-I(w)]}.
\ee
With Eq.~(\ref{Iw}) one then obtains
\be
\label{ftexp}
I(-w)-I(w)=\frac{w}{m\tau_m}+a\Delta,
\ee
with the convention $w>0$, so that the fluctuation relation is given by
\be
\label{dryft}
\frac{P_\tau(w)}{P_\tau(-w)}\approx e^{\tau\gamma\left[\frac{w}{m\tau_m}+a\Delta\right]}.
\ee
This fluctuation relation differs from the SSFT, Eq.~(\ref{conventional}), by the term $\tau\gamma a\Delta$ in the exponent, which scales linearly with the dry friction coefficient $\Delta$ and leads to a discontinuity at $w=0$ of magnitude $2a\Delta$ (cf.~Fig.~\ref{Fig_dryft}). Without dry friction, i.e., for $\Delta=0$, one recovers the conventional fluctuation relation from Eq.~(\ref{dryft}) with $c$ given by
\be
c=\frac{\gamma}{m\tau_m}.
\ee
This is expected, since for $\Delta=0$ our model Eq.~(\ref{model}) maps formally onto the paradigmatic nonequilibrium particle model, investigated by van Zon and Cohen \cite{VanZon03}, for which the conventional work fluctuation theorem was shown to hold. Note, however, that in the latter model the Gaussian noise originates from an equilibrium heat bath leading to the constant $c=1/k_BT$. Note also that, if $P(W_\tau)$ were to be continuous because of added corrections coming from the Jacobian term in the path integral, then there would be no discontinuity in Eq.~(\ref{ftexp}) or in Fig.~\ref{Fig_dryft}, as the added term $a\Delta$ would simply disappear.

\section{Concluding remarks}

\begin{enumerate}
\item We have investigated a simple phenomenological model Eq.~(\ref{model}) for the motion of a solid object, which moves over a randomly vibrating solid plate and is subject to a constant external force. The most probable or optimal path between an initial state $(v_0,t_0)$ and a final state $(v,t)$ in the $(v,s)$-plane can be a direct or an indirect path, which correspond, respectively, to a slip  motion with $v\neq 0$ or to a stick-slip motion, where the object is stuck on the plate for some time. These two kinds of motion underlie the behavior of many, more complicated systems with solid-solid friction. From a mathematical point of view, our model also represents one of the simplest models for which singular features of optimal paths can be investigated analytically (see also \cite{Dykman1994}).

\item To complement our study of the random motion of the solid object, we have investigated the fluctuations of the mechanical work performed on the object by the external force. Here, the path integral approach allows a straightforward derivation of the rate function for the stationary work distribution, which appears to be characterized by a discontinuity at zero work due to the singularity of dry friction at $v=0$. The rate function was used to formulate a fluctuation relation for the work fluctuations by taking the ratio of the probability of observing positive work values to that of negative work values. The resulting fluctuation relation also exhibits a discontinuity at zero work, while behaving linearly with $w$ away from zero, similarly to the SSFT.

\item In addition to the mechanical work due to the constant force, which is given by Eq.~(\ref{work}), one could also consider the work done on the object by the vibrations of the plate due to the externally imposed noise. Over a time $\tau$, the amount of work performed due to these vibrations is
\be
W_\tau^{vib}=\int_0^\tau v(s)\xi(s)\D s.
\ee
An investigation of the fluctuation properties of this quantity requires considerable analytical efforts (cf.~\cite{Farago02} for the case of Brownian motion), which is beyond the scope of this paper.

\item In order to be able to compare our analytical expressions with results of numerical simulations, one has to consider corrections for finite $\gamma$, since simulations are performed with finite noise strength, while our theory is strictly only valid for $\gamma\ra\infty$. One correction term is given by the contribution of the Jacobian $J[v(s)]$ in the path integral Eq.~(\ref{path_int}), which is explicitly given by Eq.~(\ref{jacobian}). This term gives a $\gamma$-independent prefactor in the saddlepoint approximation Eq.~(\ref{sp_approx}) and can only be neglected in the limit $\gamma\ra\infty$. Another correction term arises from the expansion of the action in the neighborhood of the optimal path $v^*$ leading to the so-called fluctuation factor $F[v^*]$ \cite{Hunt81,Kleinert}. The saddlepoint approximation including these two correction terms is then \cite{Baule10}
\be
f(v,t|v_0,t_0)\cong J[v^*]e^{-\gamma A[\dot{v}^*,v^*]}F[v^*],
\ee
where $J[v^*]$ is given by Eq.~(\ref{jacobian}) and the fluctuation factor by
\be
\label{fluc_fac}
F[v^*]&=&\int_{(0,t_0)}^{(0,t)}e^{-\gamma\int_{t_0}^t\D s\left[\dot{z}(s)^2+\Omega(v^*(s))z(s)^2\right]}\mathcal{D}z(s),
\ee
with $\Omega(v)= U''(v)^2+U'(v)U'''(v)$. Both correction terms contain delta-function singularities and require careful treatment. In order to avoid ambiguities arising from these singularities, one might consider a regularization of the $\sigma(v)$ term in Eq.~(\ref{model}), e.g., by replacing $\sigma(v)$ by $\tanh(v/\epsilon)$, where $\epsilon$ is a small parameter. This way one obtains a nonlinear but not singular equation of motion for the object. The singular behavior is then recovered by taking the limit $\epsilon\ra 0$.\footnote{This is reminiscent of taking the thermodynamic limit for a system in thermal equilibrium, which is necessary to understand a discontinuous first-order phase transition. In this context, one could also consider the discontinuity in the slope of the transition probability $f(v,t|v_0,t_0)$ as analogous to a first-order phase transition in equilibrium. The action would then be analogous to the equilibrium free energy.}

\item Dry friction depends on the properties of the interface between the two three-dimensional solids and is thus an intrinsic two-dimensional effect. It would be interesting to study the behavior of two-dimensional generalizations of the one-dimensional model given by Eq.~(\ref{model}), possibly with a non-uniform surface roughness, and to compare the results of these models with experiments, in order to obtain information about surface properties, such as roughness, defects, etc.

\item Further extensions of the phenomenological model Eq.~(\ref{model}) are possible, by adding, e.g., an external linear spring force acting on the object. Models of this type, in the absence of external vibrations, have been considered as a common type of model in the literature on dry friction \cite{Elmer97,Urbakh04}. Another possible extension concerns the statistics of the externally imposed vibrations. By considering, e.g., Poissonian shot noise \cite{Feynman} (see also \cite{Baule09}), which allows, in principle, a great variety of pulse characteristics absent in Gaussian noise, one might obtain new effects in the stick-slip motion of the object.

\item The linear dynamic friction $-\alpha v(t)$ in Eq.~(\ref{model}) is present for a number of solid-solid systems for large velocities or for solid-fluid systems (lubricated interfaces). For smaller velocities, a different and more complicated nonlinear behavior is often observed, due to the nature of the object-solid interface \cite{Baumberger96}. Without going into the meso- or microscopic nature of the velocity dependence of these measured dynamical friction coefficients, it would be instructive to investigate the phenomenological model Eq.~(\ref{model}), using a simple mathematical representation of the measured dynamical friction as a function of the object's velocity.

In carrying this out, one would study an equation with two different kinds of nonlinearities, associated with both dynamical and dry friction. It would be interesting to study the physical difference of the system's behavior due to these two nonlinearities.

\item A recent experimental investigation of the motion of a solid object moving over a rough plate subject to externally-imposed, random vibrations and a constant external force (gravity), as described by Eq.~(\ref{model}), has been performed by Chaudhury \textit{et al} \cite{Goohpattader09,Goohpattader10}. Here, the trajectories of a small glass prism, which moves over an inclined glass plate in the presence of external white Gaussian noise and gravity, are recorded using a high-speed video camera. This investigation focuses mainly on the properties of the drift velocity as well as the diffusive behavior of the prism, where the diffusion coefficient is estimated from the variance of the probability distribution of the prism's displacement. It would be interesting to compare a theoretical prediction for the diffusion coefficient of the prism, following from our theory, with the experimental results obtained in \cite{Goohpattader09,Goohpattader10}. The diffusion coefficient $D_p$ of the prism could be evaluated using the Green-Kubo relation (cf. \cite{deGennes05})
\be
D_p&=&\int_0^\infty\left<v(0)v(t)\right>\D t.
\ee
The velocity correlation function $\left<v(0)v(t)\right>$ is then obtained from the transition probability $f(v,t|v_0,t_0)$ by
\be
\left<v(0)v(t)\right>=\int_0^\infty\D v \int_0^\infty\D v_0\, v\,v_0\,f(v,t|v_0,0)p(v_0),
\ee
where $p(v_0)$ is the stationary velocity distribution Eq.~(\ref{p_stat}).

In \cite{Goohpattader10}, the fluctuations of the mechanical work rate, Eq.~(\ref{work}), performed by the constant external force on the prism have also been determined. In fact, a fluctuation relation that increases linearly with the work rate has been measured, which would be in partial agreement with our result Eq.~(\ref{dryft}). However, the jump at zero work rate in the fluctuation relation, as predicted by Eq.~(\ref{dryft}), was not observed. A more detailed comparison of our theory with these experiments is left for the future.

\item Another interesting experimental test of our theory could be performed in an electrical circuit, based on a well-known mapping of a class of mechanical systems on electrical circuits \cite{VanZon04b,Taniguchi08b}. In that case, the dry friction can be represented by a diode (cf. \cite{Elmer97}). Electric circuits are particularly suitable for investigating the properties and physical significance of optimal paths (see, e.g., \cite{Dykman1992,Dykman1994,Luchinsky1998}).

\item It might be interesting to investigate our model in the context of the recent transition path theory \cite{Eijnden2006,Metzner2006}. Open questions concern the nature of the singularities in the transition paths and their connection with optimal paths in the path integral framework.
\end{enumerate}

\begin{acknowledgments}
HT is supported by RCUK (Interdisciplinary Fellowship).
\end{acknowledgments}


\begin{thebibliography}{12}
\expandafter\ifx\csname natexlab\endcsname\relax\def\natexlab#1{#1}\fi
\expandafter\ifx\csname bibnamefont\endcsname\relax
  \def\bibnamefont#1{#1}\fi
\expandafter\ifx\csname bibfnamefont\endcsname\relax
  \def\bibfnamefont#1{#1}\fi
\expandafter\ifx\csname citenamefont\endcsname\relax
  \def\citenamefont#1{#1}\fi
\expandafter\ifx\csname url\endcsname\relax
  \def\url#1{\texttt{#1}}\fi
\expandafter\ifx\csname urlprefix\endcsname\relax\def\urlprefix{URL }\fi
\providecommand{\bibinfo}[2]{#2}
\providecommand{\eprint}[2][]{\url{#2}}

\bibitem{Baule10} A. Baule, E. G. D. Cohen, and H. Touchette, J. Phys. A \textbf{43}, 025003 (2010).

\bibitem[{\citenamefont{Persson}(2000)}]{Persson}
\bibinfo{author}{\bibfnamefont{B.~N.~J.} \bibnamefont{Persson}},
  \emph{\bibinfo{title}{Sliding Friction: Physical Principles and Applications}} (\bibinfo{publisher}{Springer},
  \bibinfo{year}{2000}).

\bibitem[{\citenamefont{Gennes}(2005/06/01/)}]{deGennes05}
\bibinfo{author}{\bibfnamefont{P.-G.} \bibnamefont{de Gennes}},
  \bibinfo{journal}{J. Stat. Phys.}
  \textbf{\bibinfo{volume}{119}}, \bibinfo{pages}{953}
  (\bibinfo{year}{2005}).

\bibitem{Hayakawa05} H. Hayakawa, Physica D \textbf{205}, 48 (2005).

\bibitem{Risken} H. Risken, \textit{The Fokker-Planck Equation: Methods of Solution and Applications} (Springer, 1996).

\bibitem[{\citenamefont{Feynman and Hibbs}(1965)}]{Feynman}
\bibinfo{author}{\bibfnamefont{R.~P.} \bibnamefont{Feynman}} \bibnamefont{and}
  \bibinfo{author}{\bibfnamefont{A.~R.} \bibnamefont{Hibbs}},
  \emph{\bibinfo{title}{{Quantum Mechanics and Path Integrals}}}
  (\bibinfo{publisher}{McGraw-Hill, New York}, \bibinfo{year}{1965}).

\bibitem[{\citenamefont{Onsager and Machlup}(1953)}]{Onsager53}
\bibinfo{author}{\bibfnamefont{L.}~\bibnamefont{Onsager}} \bibnamefont{and}
  \bibinfo{author}{\bibfnamefont{S.}~\bibnamefont{Machlup}},
  \bibinfo{journal}{Phys. Rev.} \textbf{\bibinfo{volume}{91}},
  \bibinfo{pages}{1505} (\bibinfo{year}{1953}).

\bibitem[{\citenamefont{Machlup and Onsager}(1953)}]{Machlup53}
\bibinfo{author}{\bibfnamefont{S.}~\bibnamefont{Machlup}} \bibnamefont{and}
  \bibinfo{author}{\bibfnamefont{L.}~\bibnamefont{Onsager}},
  \bibinfo{journal}{Phys. Rev.} \textbf{\bibinfo{volume}{91}},
  \bibinfo{pages}{1512} (\bibinfo{year}{1953}).

\bibitem[{\citenamefont{Graham}(1973)}]{Graham73}
\bibinfo{author}{\bibfnamefont{R.}~\bibnamefont{Graham}},
  \bibinfo{journal}{Springer Tracts in Modern Physics}
  \textbf{\bibinfo{volume}{66}}, \bibinfo{pages}{1} (\bibinfo{year}{1973}).

\bibitem[{\citenamefont{Hunt and Ross}(1981)}]{Hunt81}
\bibinfo{author}{\bibfnamefont{K.~L.~C.} \bibnamefont{Hunt}} \bibnamefont{and}
  \bibinfo{author}{\bibfnamefont{J.}~\bibnamefont{Ross}}, \bibinfo{journal}{J. Chem. Phys.} \textbf{\bibinfo{volume}{75}},
  \bibinfo{pages}{976} (\bibinfo{year}{1981}).
  
\bibitem{Freidlin84} M. I. Freidlin and A. D. Wentzell, \textit{Random Perturbations of Dynamical Systems} (Springer, 1984).

\bibitem{Touchette09b} H. Touchette, Phys. Rep. \textbf{478}, 1 (2009).

\bibitem[{\citenamefont{Taniguchi and Cohen}(2007)}]{Taniguchi07}
\bibinfo{author}{\bibfnamefont{T.}~\bibnamefont{Taniguchi}} \bibnamefont{and}
  \bibinfo{author}{\bibfnamefont{E.~G.~D.} \bibnamefont{Cohen}},
  \bibinfo{journal}{J. Stat. Phys.}
  \textbf{\bibinfo{volume}{126}}, \bibinfo{pages}{1} (\bibinfo{year}{2007}).

\bibitem[{\citenamefont{Harris and Schutz}(2007)}]{Harris07}
\bibinfo{author}{\bibfnamefont{R.~J.} \bibnamefont{Harris}} \bibnamefont{and}
  \bibinfo{author}{\bibfnamefont{G.~M.} \bibnamefont{Schutz}},
  \bibinfo{journal}{J. Stat. Mech.}, \bibinfo{pages}{P07020}
  (\bibinfo{year}{2007}).

\bibitem[{\citenamefont{Evans et~al.}(1993)\citenamefont{Evans, Cohen, and
  Morriss}}]{EvansD93}
\bibinfo{author}{\bibfnamefont{D.~J.} \bibnamefont{Evans}},
  \bibinfo{author}{\bibfnamefont{E.~G.~D.} \bibnamefont{Cohen}},
  \bibnamefont{and} \bibinfo{author}{\bibfnamefont{G.~P.}
  \bibnamefont{Morriss}}, \bibinfo{journal}{Phys. Rev. Lett.}
  \textbf{\bibinfo{volume}{71}}, \bibinfo{pages}{2401} (\bibinfo{year}{1993}).

\bibitem[{\citenamefont{Gallavotti and Cohen}(1995)}]{Gallavotti95}
\bibinfo{author}{\bibfnamefont{G.}~\bibnamefont{Gallavotti}} \bibnamefont{and}
  \bibinfo{author}{\bibfnamefont{E.~G.~D.} \bibnamefont{Cohen}},
  \bibinfo{journal}{Phys. Rev. Lett.} \textbf{\bibinfo{volume}{74}},
  \bibinfo{pages}{2694} (\bibinfo{year}{1995}).

\bibitem[{\citenamefont{Kurchan}(1998)}]{Kurchan98}
\bibinfo{author}{\bibfnamefont{J.}~\bibnamefont{Kurchan}},
  \bibinfo{journal}{J. Phys. A: Math. Gen.}
  \textbf{\bibinfo{volume}{31}}, \bibinfo{pages}{3719} (\bibinfo{year}{1998}).
  
  \bibitem[{\citenamefont{Lebowitz and Spohn}(1999)}]{Lebowitz99}
\bibinfo{author}{\bibfnamefont{J.~L.} \bibnamefont{Lebowitz}} \bibnamefont{and}
  \bibinfo{author}{\bibfnamefont{H.}~\bibnamefont{Spohn}},
  \bibinfo{journal}{J. Stat. Phys.}
  \textbf{\bibinfo{volume}{95}}, \bibinfo{pages}{333} (\bibinfo{year}{1999}).

\bibitem[{\citenamefont{{van Zon} and Cohen}(2003{\natexlab{a}})}]{VanZon03}
\bibinfo{author}{\bibfnamefont{R.}~\bibnamefont{{van Zon}}} \bibnamefont{and}
  \bibinfo{author}{\bibfnamefont{E.~G.~D.} \bibnamefont{Cohen}},
  \bibinfo{journal}{Phys. Rev. E} \textbf{\bibinfo{volume}{67}},
  \bibinfo{pages}{046102} (\bibinfo{year}{2003}{\natexlab{a}}).

\bibitem{Dykman1994} M. I. Dykman, M. M. Millonas, and V. N. Smelyanskiy, Phys. Lett. A \textbf{195}, 53 (1994).

\bibitem{Farago02} J. Farago, J. Stat. Phys. \textbf{107}, 781 (2002).

\bibitem{Kleinert} H. Kleinert, \textit{Path Integrals in Quantum Mechanics, Statistics, Polymer Physics, and Financial Markets} (World Scientific, 2009).

\bibitem{Elmer97} F.-J. Elmer, J. Phys. A: Math Gen \textbf{30}, 6057 (1997).

\bibitem{Urbakh04} M. Urbakh, J. Klafter, D. Gourdon, and J. Israelachvili, Nature \textbf{430}, 525 (2004).

\bibitem[{\citenamefont{Baule and Cohen}(2009{\natexlab{a}})}]{Baule09}
\bibinfo{author}{\bibfnamefont{A.}~\bibnamefont{Baule}} \bibnamefont{and}
  \bibinfo{author}{\bibfnamefont{E.~G.~D.} \bibnamefont{Cohen}},
  \bibinfo{journal}{Phys. Rev. E} \textbf{\bibinfo{volume}{79}}, \bibinfo{eid}{030103(R)}
  (\bibinfo{year}{2009}{\natexlab{a}}); \textit{ibid.}, \bibinfo{journal}{Phys. Rev. E} \textbf{\bibinfo{volume}{80}}, \bibinfo{eid}{011110}
   (\bibinfo{year}{2009}{\natexlab{b}}).
   
\bibitem{Baumberger96} T. Baumberger, \textit{Dry friction dynamics at low velocities} in \textit{Physics of Sliding Friction}, edited by B. N. J. Persson and E. Tosatti, (Nato ASI Series, Springer, 1996).

\bibitem{Goohpattader09} P. S. Goohpattader, S. Mettu, and M. K. Chaudhury, Langmuir \textbf{25}, 9969 (2009).

\bibitem{Goohpattader10} P. S. Goohpattader and M. K. Chaudhury, J. Chem. Phys. \textbf{133}, 024702 (2010).

\bibitem{VanZon04b} R. Van Zon, E. G. D. Cohen, and S. Ciliberto, Phys. Rev. Lett. \textbf{92}, 130601 (2004).

\bibitem{Taniguchi08b} T. Taniguchi and E. G. D. Cohen, J. Stat. Phys. \textbf{130}, 633 (2008).

\bibitem{Dykman1992} M. I. Dykman, P. V. E. McClintock, V. N. Smelyanski, N. D. Stein, and N. G. Stocks, Phys. Rev. Lett. \textbf{68}, 2718 (1992).

\bibitem{Luchinsky1998} D. G. Luchinsky, P. V. E. McClintock, and M. I. Dykman, Rep. Prog. Phys. \textbf{61}, 889 (1998).

\bibitem{Eijnden2006} W. E. and E. Vanden-Eijnden, J. Stat. Phys. \textbf{123}, 503 (2006)

\bibitem{Metzner2006} P. Metzner, C. Sch\"utte, and E. Vanden-Eijnden, J. Chem. Phys. \textbf{125}, 084110 (2006).

\end{thebibliography}
\end{document}